\documentclass[11pt]{abpaper}
\usepackage{xcolor,tikz,graphicx}
\usepackage{todonotes}
\usepackage{hyperref}
\usepackage{xspace}
\usepackage{relsize}
\usepackage[update,prepend]{epstopdf}
\usepackage{listings}
\usepackage{amsmath,amssymb}
\usepackage{cancel}
\usepackage{url}
\usepackage{microtype}
\usepackage{booktabs}
\usepackage{bm}
\usepackage{verbatim}
\usepackage{multirow}

\newcommand{\eV}{\ensuremath{\text{e}\mspace{-0.8mu}\text{V}\xspace}}

\newcommand{\GeV}{\ensuremath{\text{G\eV}}\xspace}

\newcommand{\mb}{\ensuremath{\text{mb}}\xspace}
\newcommand{\nb}{\ensuremath{\text{nb}}\xspace}

\newcommand{\pt}{\ensuremath{p_T}\xspace}

\newcommand{\alphaS}{\ensuremath{\alpha_\mathrm{S}}\xspace}
\newcommand{\alphaQED}{\ensuremath{\alpha_\mathrm{QED}}\xspace}
\newcommand{\dr}{\ensuremath{\Delta{R}}\xspace}

\newcommand{\kt}{\ensuremath{k_T}\xspace}
\newcommand{\CA}{\ensuremath{\text{C/\!A}}\xspace}
\newcommand{\akt}{\ensuremath{\text{anti-\kt}}\xspace}

\newcommand{\Sherpa}{\textsc{Sherpa}\xspace}

\newcommand{\Pythia}{\textsc{Pythia}\xspace}
\newcommand{\Herwig}{\textsc{Herwig}\xspace}

\newcommand{\xx}{\raisebox{0.25ex}{\smaller[2]++}\xspace}

\newcommand{\img}[2][1.0]{\includegraphics[width=#1\textwidth]{#2}}

\makeatletter
\DeclareRobustCommand{\kbd}[1]{{\smaller{\texttt{#1}}}}

\g@addto@macro\bfseries{\boldmath}
\makeatother

\usepackage{mathpazo}

\usepackage{fancyref}
\usepackage{boxedminipage}
\usepackage{placeins}

\author{\textbf{Andy Buckley, Chris Pollard}\\[0.1mm] \rmfamily\smaller\emph{School of Physics \& Astronomy, Glasgow University, UK}}
\title{QCD-aware partonic jet clustering for truth-jet flavour labelling}
\preprints{MCnet-15-12\\ GLA-PPE-2015-03}

\begin{document}

\begin{abstract}
  We present an algorithm for deriving partonic flavour labels to be applied to
  truth particle jets in Monte Carlo event simulations. The inputs to this
  approach are final pre-hadronisation partons, to remove dependence on
  unphysical details such as the order of matrix element calculation and shower
  generator frame recoil treatment. These are clustered using standard jet
  algorithms, modified to restrict the allowed pseudojet combinations to those
  in which tracked flavour labels are consistent with QCD and QED Feynman
  rules. The resulting algorithm is shown to be portable between the major
  families of shower generators, and largely insensitive to many possible
  systematic variations: it hence offers significant advantages over existing
  \textit{ad hoc} labelling schemes. However, it is shown that contamination
  from multi-parton scattering simulations can disrupt the labelling results.
  Suggestions are made for further extension to incorporate more detailed
  QCD splitting function kinematics,
  robustness improvements, and potential uses for truth-level physics object
  definitions and tagging.
\end{abstract}


\let\OLDsection*\section*
\renewcommand


\section*{Introduction}

The rise of jet substructure methods at the LHC has prompted a resurgence in
attempts to distinguish ``quark'' and ``gluon'' hadronic jets from each other,
primarily for use in searches for BSM phenomena. Such attempts are primarily
based on the different colour charges of quarks and gluons, with the larger
$C_A$ colour factor of the gluon associated with more jet broadening and higher
constituent multiplicities than the quarks' $C_F$ factor.

A conceptual problem immediately arises in that colour-neutral jets cannot
perfectly correspond to coloured single partons. Additionally, final-state
observables do not provide unambiguous evidence for two distinct statistical
populations of hadronic jets. The evaluation of $q/g$ jet tagging methods has
hence been based on use of event generators' partonic event graphs to define the
``true'' jet label, typically assigning a partonic identity to each final-state
truth-jet by the flavour of the highest-energy or highest-\pt parton found
within a fixed angular distance of its centroid. Sometimes these labelling
partons are chosen from the entire partonic event record, including parton
shower (PS) evolution; sometimes they are restricted to partons associated to
the hard process matrix element (ME). The labels derived using these methods are
then used to determine the efficiencies and fake rates of experimental tagging
techniques. Such approaches to truth labelling have several practical and
conceptual drawbacks:
\begin{itemize}
\item their reliance on the unguaranteed physicality of partonic event records
  -- these are typically intended more for debugging than for robust physics
  analysis use, and their momenta may not have physical meaning;
\item even where the parton momenta are physical in their chosen frame, it is
  often the case that different generations of parton evolution are represented
  in frames different from the lab frame relevant to final state observables;
\item the potential unphysical distinction between ``matrix element'' and
  ``parton shower'' partons -- problematic for consistency of labelling at
  different perturbative orders and particularly for ``resummation-corrected''
  matrix elements such as those in the POWHEG method~\cite{Frixione:2007vw} where
  there is no clear kinematic distinction between ME and PS emissions.
\end{itemize}

All of these limitations and assumptions cause problems in practice, notably in
the inability to use the \Sherpa event generator~\cite{Gleisberg:2008ta} whose event
record is complexified by the use of matrix-element/parton-shower merging and
matching (MEPS) and a dipole shower formalism with $2 \to 3$ parton
branchings~\cite{Schumann:2007mg}.  More traditional parton showers, i.e. the $1 \to
2$ parton splitting formulations used in the \Pythia~\cite{pythia6,Sjostrand:2007gs,Sjostrand:2014zea} and
\Herwig generator families~\cite{fherwig,Bahr:2008pv}, are themselves problematic
due to the need for momentum reshuffling to preserve Lorentz invariance in PS
evolution: their off-shell intermediate parton representations may not have
physically reasonable four-momenta, nor even be represented in a well-defined
reference frame. Practically, even if physical information could in principle be
obtained for every parton in a given MC generator's event record, it would be
extremely inconvenient to require separate algorithms for each generator's event
records.

In this paper we propose and characterise an alternative truth-jet labelling
method, based on standard jet clustering algorithms and measures, modified to
only permit clusterings compatible with $1 \to 2$ QCD and QED processes. This
has been implemented in the FastJet framework for three standard jet measures,
and we present studies of the performance, robustness and portability of the
resulting labels across physics processes, perturbative orders, PS event
generators, distance measures, and soft physics effects.

The primary motivation for this work is to provide an operational definition
without the practical problems of ill-defined clustering inputs and
generator-incompatibility which plague existing truth labelling
schemes. However, we note the close relation of this scheme to prior work on
infrared safety in partonic jet definition~\cite{Banfi:2006hf} and discuss
possibilities for extension of this scheme both for improved perturbative safety
and for use in tagging beyond the regime of Sudakov emission kinematics.

\section{QCD-aware parton clustering}


The ``QCD-aware'' clustering algorithm which we will characterise in this paper
is directly intended to address the limitations enumerated in the previous
section. Our priority is to obtain a robust truth-jet labelling scheme usable
with all event generator codes, and based on physically well-defined partons,
and not at present the issues of QCD singularities focused on by the related
flavour-\kt algorithm~\cite{Banfi:2006hf}.

\subsection{Jet clustering algorithms}

QCD-aware clustering is, like flavour-\kt, a modification of the
well-established $\kt^{2n}$ family of agglomerative jet clusterings, developed
as an infrared \& collinear safe alternative to cone-based jet
finding~\cite{Salam:2009jx}. These operate by clustering an initial population
of would-be jet constituents (particles, or experimental objects such as
calorimeter cell clusters), known as \textit{pseudojets}, into final jets by
repeated $2 \to 1$ combinations of pseudojets. At each iteration of the
algorithm, the pair of pseudojets to be combined is that which minimises a
distance measure containing energy and angular terms. The distance measure for
the $\kt^{2n}$ family is
\begin{equation}
  \label{eq:distkt2n}
  d^{(n)}_{ij} = \min{\left({\kt^{2n}}_i, {\kt^{2n}}_j\right)} \Delta{R^2_{ij}}/R^2
\end{equation}
where $i$ and $j$ are the pseudojet indices, $\Delta{R_{ij}}$ is the
beamline-boost-invariant distance between them in $\eta$--$\phi$ or $y$--$\phi$
space, $R$ is a parameter defining the characteristic radius of the resulting
jets, and the choice of $n$ exponent chooses whether the algorithm is ``\kt''
($n=1$), ``Cambridge--Aachen (\CA)''($n=0$), or ``\akt''
($n=-1$)\footnote{Other, potentially non-integer, values could also be used but
  have received little attention since the sign of $n$ is more important than
  its absolute value.}. A pseudojet $i$ is considered ``final'' and removed from
the clustering if its nearest distance to another pseudojet is greater than
${\kt^{2n}}_i$, the so-called beam distance. Clustering stops when no pseudojets
remain.

The formal origins of this family of clustering distances lie in the form of the
measure, which for the original \kt measure is proportional to both the smallest
transverse momentum and to $\Delta{R}$. This focus on the low-\kt and small
emission angle regions ensures resummation of both the collinear and soft
divergences in the QCD gluon emission splitting function, and \kt clustering is
often referred to as an inversion of the QCD emission sequence of a parton
shower or resummation\footnote{The aim of the flavour-\kt algorithm was to make
  this inverse relationship more precise, by using a distinct distance measure
  for the purely collinear divergence of the $g \to q\bar{q}$ parton
  splitting.}. The later \CA and \akt algorithms only address the
collinear divergence, but due to its production of circular jets the latter has
become the standard jet type used at the LHC. The FastJet~\cite{Cacciari:2011ma} package
contains several important optimisations for the $\kt^{2n}$ family, making use
of geometric properties of the clustering measures to ameliorate the na\"ive
cubic dependence on the number of initial pseudojets.

\subsection{Flavour-aware clustering}

The relationship between jet clustering and dominant QCD emission kinematics is
central to the QCD-aware approach to truth jet labelling. The ``first'' partons
connected to a MC generator signal process suffer from an unphysical distinction
between matrix element and shower emissions, as well as uncertainty over the
physicality of their momenta, but inversion of the emission sequence starting
from more physical final partons should in principle be well-behaved.

\begin{figure}[t]
  \centering

  \includegraphics{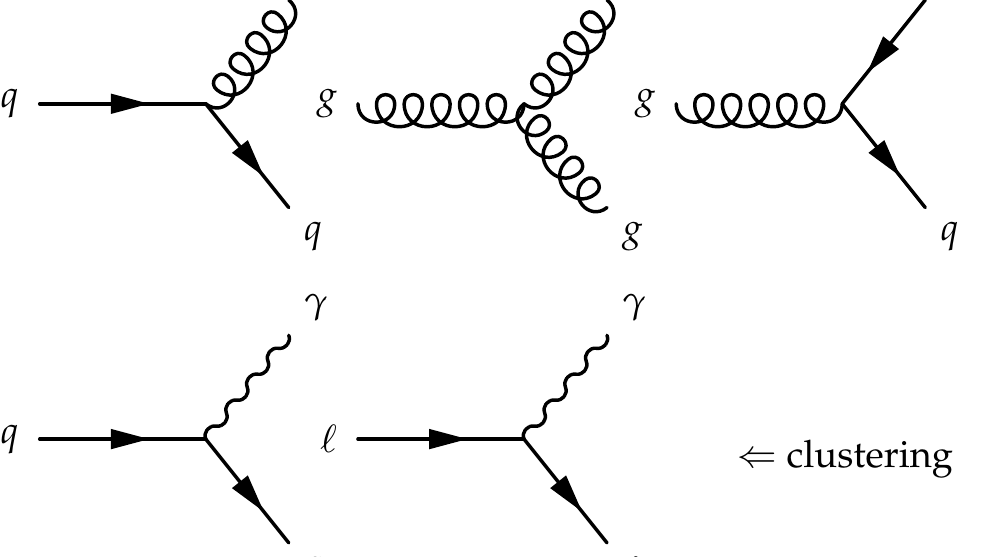}
















  \vspace*{1em}
  \caption{Feynman rule vertices used for QCD (and QED) aware jet clustering.}
  \label{fig:feynrules}
\end{figure}

We hence propose a minor modification to the above family of clustering
algorithms, in which the measure is adapted to ensure that only clusterings
compatible with the 3-point Feynman rules of QCD and QED will be granted a
finite distance. The vertices corresponding to these Feynman rules are shown in
Figure~\ref{fig:feynrules}. Note that charged leptons and photons are also
included. The ``QCD-aware'' distance measure is hence
\begin{equation}
  \label{eq:distqcda}
  D_{ij}^{(n)} =
  \begin{cases}
    d_{ij}^{(n)} & \text{if flavours QCD/QED compatible},\\
    \infty     & \text{otherwise}.
  \end{cases}
\end{equation}
As the effect is to veto flavour-incompatible clusterings which \emph{would}
take place in a standard flavour-blind jet algorithm, all the FastJet
performance optimisations for the $\kt^{(n)}$ family could also apply to
the QCD-aware variants of those algorithms.

For this to work, each pseudojet must carry an extra fundamental particle
flavour label in addition to the usual kinematic information, and this label
must be set on the resulting pseudojet from each $2 \to 1$ merging, as required
by consistency with the corresponding Feynman rule. This measure modification
and relabelling algorithm are implemented in the \kbd{QCDAware} FastJet plugin
class.




It is important to note some simplifications in our scheme. First, we have
ignored the 4-gluon vertex since it cannot be mapped to a $2 \to 1$ clustering
and has a small splitting function. Secondly, we do not permit ambiguity: a
$q\bar{q}$ clustering could in principle resolve to either a gluon (as
implemented) or a photon\footnote{For now we also do not consider the
  $\ell^+\ell^- \to \gamma$ inverted photon conversion as an allowed
  clustering.}. One could consider both ``histories'' to be valid, assign
weights according to computed relative probabilities, and eventually marginalise
labelling results across the weighted ensemble of potential histories: this
would be an approach along the lines of the shower deconstruction
method~\cite{Soper:2011cr}. Instead, for definiteness and computational
tractability, we ignore the QED solution in favour of the much stronger strong
coupling. Finally, we use a bare \kt measure where the flavours are consistent,
with no weighting for distinct splitting kinematics cf. flavour-\kt, nor for the
relative strengths of (running) \alphaS and \alphaQED or the relevant charges of
the participating particles. This last set of points is certainly worthy of
consideration, but as we shall see even the bare algorithm exhibits interesting
and useful features.




Finally we highlight a somewhat obvious property of flavour-aware clustering
which is useful to bear in mind: that the majority of emissions consist of
emission of the gauge boson, i.e. the gluon or photon for strong and EM
radiation respectively, and that this produces no flavour change. We would
expect that it would be relatively hard for a quark jet to lose its quark label
(this would require an accidental clustering with an antiquark of the same
flavour), while a true gluon jet can be mislabelled as a quark jet by the
accidental presence of a single quark within its capture radius. (The same
applies to should-be isolated photons contaminated by the proximity of a quark
or charged lepton.) We will consider this effect later, particularly when
studying systematic variations on MC generator configurations.


\section{Selection of clustering inputs}
\label{sec:inputs}


Before the "QCD-aware" clustering rules can be used to generate labelling
parton-jets, we must identify the partons, photons, and leptons to be clustered.

Only the final quarks and gluons --- those immediately before hadronisation ---
are used as inputs to the clustering. This avoids double-counting, since the
partons in question are guaranteed not to have undergone any further splittings or
radiation, and we have verified that the three main families of parton shower MC
generators record the momenta of these partons in the lab frame\footnote{This is
  not a standard imposed by the HepMC~\cite{Dobbs:2001ck} event record standard, but
  perhaps it should be.}.
The representation of final partons is not uniform between generators, so we
apply a two-step heuristic to identify them: first we accept a parton if it is
incoming to a vertex with $\text{status/id} = 5$, the standard code for a
hadronisation vertex; and secondly, if the first condition is not met, the
particle is accepted if it has no children which are also partons. The first
condition is preferable, but hadronisation vertex labelling is currently only
implemented in Sherpa, hence the second heuristic is required to identify final
partons in Herwig\xx and Pythia\,8.

Photons, electrons, muons, and hadronically-decaying taus that have no hadronic
ancestors (``prompt'') are included in the clustering inputs.  The ``prompt''
restriction is because the four-momentum associated with non-prompt photons and
leptons is already included in the momenta of the final partons whose descendent
hadrons decay to them. The restriction to hadronic taus is for alignment with
experimental procedure: taus that decay to leptons are generally reconstructed
by experiments as charged leptons and missing energy rather than as jets, with
such charged leptons from prompt taus being themselves classified as prompt.

An implementation of this input-selection algorithm is available in the
\kbd{FinalPartons} and \kbd{TauFinder} projections in
Rivet~\cite{Buckley:2010ar} from version~2.2.1 onwards.

\section{Associating labels to jets}


After applying the QCD-aware clustering to the partonic (and prompt lepton and
photon) inputs discussed above, and normal flavour-blind clustering to the final
state truth particles, we have two distinct jet collections: flavoured partonic
label-jets, and standard particle jets. We aim to label the latter using the
former.

Arguably the simplest labelling algorithm is to assign each particle jet the
label of the closest parton jet, i.e. that which minimises
$\Delta{R}_\text{jet--label}$.  This has the drawback, however, that distinct
particle jets can share the same closest parton jet: should the particle jets
share the same label, or should some additional matching criterion be introduced
to assign the parton jet to just one particle jet, e.g. the nearer of the two?

In this study we have hence used the ghost association~\cite{Cacciari:2007fd}
method to non-invasively cluster the parton jets into the particle jets,
guaranteeing that no parton label will be associated to multiple particle
jets\footnote{Note that this is for the purposes of definiteness more than
  absolute physical correctness: in such ambiguous circumstances there is no
  guarantee that ghost association has picked the physically ``correct''
  assignment, or that such a thing even exists.}.

A second ambiguity now arises, because more than one parton jet can be
ghost-associated to a given particle jet. Since the QCD-aware clustering forbids
combination of some parton flavours, having multiple unclustered partons within
a particle jet's clustering radius is a fairly frequent occurrence -- moreso
than the many-particle-to-one-parton ambiguity that ghost association resolves.
Hence a disambiguation measure is required among the associated parton jets, and
for simplicity we have chosen the label which minimises
$\Delta{R}_\text{jet--label}$, within an inner core of the jet radius: if all
$\Delta{R} > 0.2$, the jet remains unlabelled. This restriction to $\Delta{R} <
0.2$ was added to remove long, low rate tails observed in initial runs of the
algorithm. This may certainly be improved, and we suggest either a combined
$\Delta{R}$ and \pt matching measure (although this is a little like adding
apples and oranges), to favour high-\pt or well-matched \pt labels within the
jet cone, and/or to assign weights rather than absolute labels -- but we do not
consider such extensions in this paper.

\section{Performance of QCD-aware labelling}


In this section we will make performance comparisons of the above-described
labelling algorithm for two hard processes, dijet and $\gamma$-jet, with various
parton shower event generators, and with several systematic variations to both
the labelling scheme and to the simulation:
\begin{description}
\item[Shower generators:] Pythia~8.201, Herwig\xx~2.7.1, Sherpa~2.1.1
\item[Clustering variants:] max-\pt (no clustering) / \kt QCD-aware / \akt QCD-aware
\item[Simulation variants:] normal / without MPI / raised shower cutoff / ME max multiplicity
\item[Association variants:] all labels / reclustered labels
\end{description}

\noindent
A minimum \pt requirement of 25\;\GeV has been imposed on the particle jets in
these studies, and a $\pt > 5\;\GeV$ requirement on the partonic label
jets. Both types of jets were clustered with an $R$ parameter of 0.4.
All jet clustering was restricted to within $|\eta| < 2.5$.







For comparison to the QCD-aware approach, we will present a ``maximum \pt''
partonic jet labelling scheme, where the label assigned to a final-state truth
jet is the flavour of the highest-\pt parton within its radius. This label is
discovered by looping over \emph{all} partons, including those in the hard
process final state (typically in the partonic centre-of-mass frame of the
matrix element calculation), through all the intermediate stages of the parton
shower and MPI, down to the final partons described in \Fref{sec:inputs}. Since
the highest-\pt parton is used, this tends to be from the hard process or
shortly after, before it has radiated significant virtuality via shower
branchings. The measures that we use for labelling kinematic performance would
be biased in this scheme, hence we will only show it in direct comparisons
either of label assignments or in ratios of flavour label rates.

\subsection{MC generator families}
\label{sec:mcgens}

A key motivation for the QCD-aware approach to partonic truth-jet labelling is
for the method to be portable between different MC generators. Each plot in most
of the following studies is hence shown with three MC lines, for the three major
parton shower MC generator families; the exact versions are given above.

In principle, the QCD-aware method should be robust enough for use both with
fixed-order codes (producing a few-body partonic final state) and parton shower
codes in which the final-state partonic multiplicity is much higher. Substantial
differences between the Herwig and Pythia generator families have been seen in
$q/g$ rate prediction studies using the max-\pt labelling
scheme~\cite{Adams:2015hiv,Larkoski:2014pca}, so some level
of variation is to be expected between generator shower formalisms, but we
expect broad qualitative agreement of labelling both between generators and
with the expectations for each hard process type.

In particular we are interested to see how the Sherpa generator compares to the
Herwig and Pythia families, since it has not previously been possible to include
Sherpa in partonic labelling studies.

In all the plots that follow in this section, all three generators have been
configured to use only lowest-order $2 \to 2$ hard scattering matrix elements,
so that any differences are due to MC family differences in parton shower
algorithms, and matrix element scale choices.

More specific modelling variations within generator families are considered in
Sections~\ref{sec:psmpisysts} and \ref{sec:mesysts}, using only Pythia and
only Sherpa respectively.

\subsection{Performance of default QCD-aware labelling}
\label{sec:qcdaperf}

\begin{figure}[p]
  \centering
  \img[0.48]{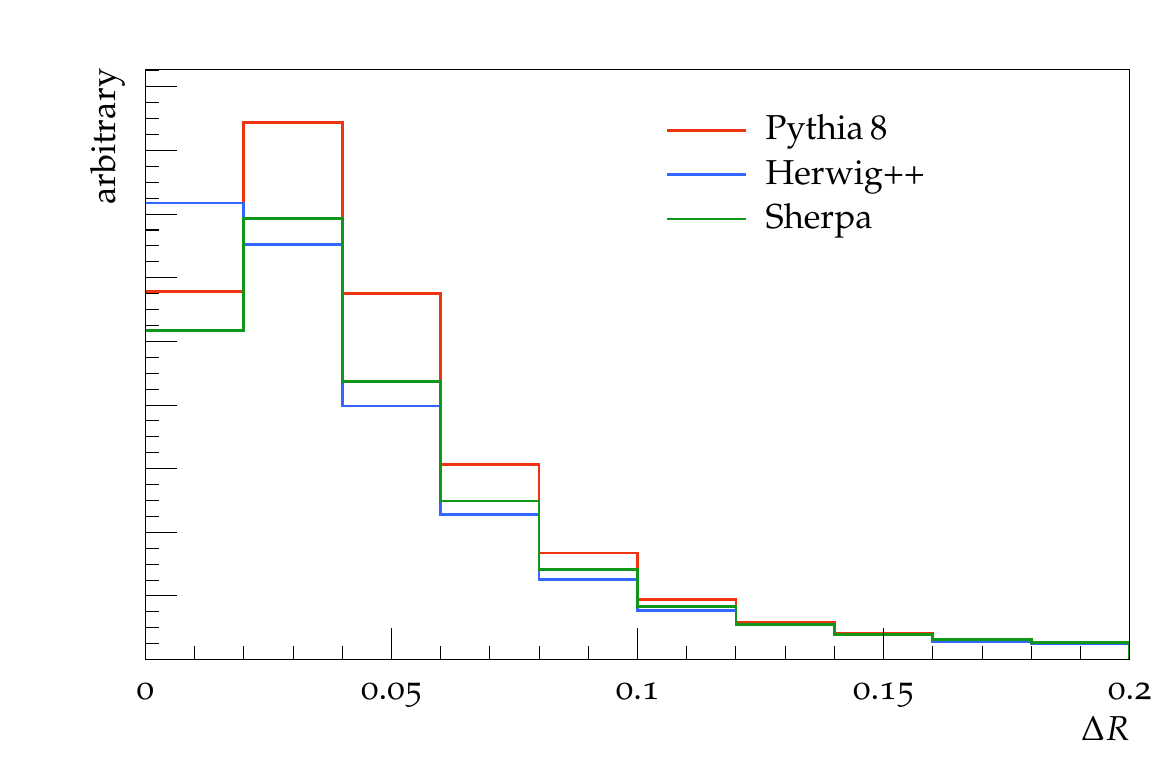} \quad \img[0.48]{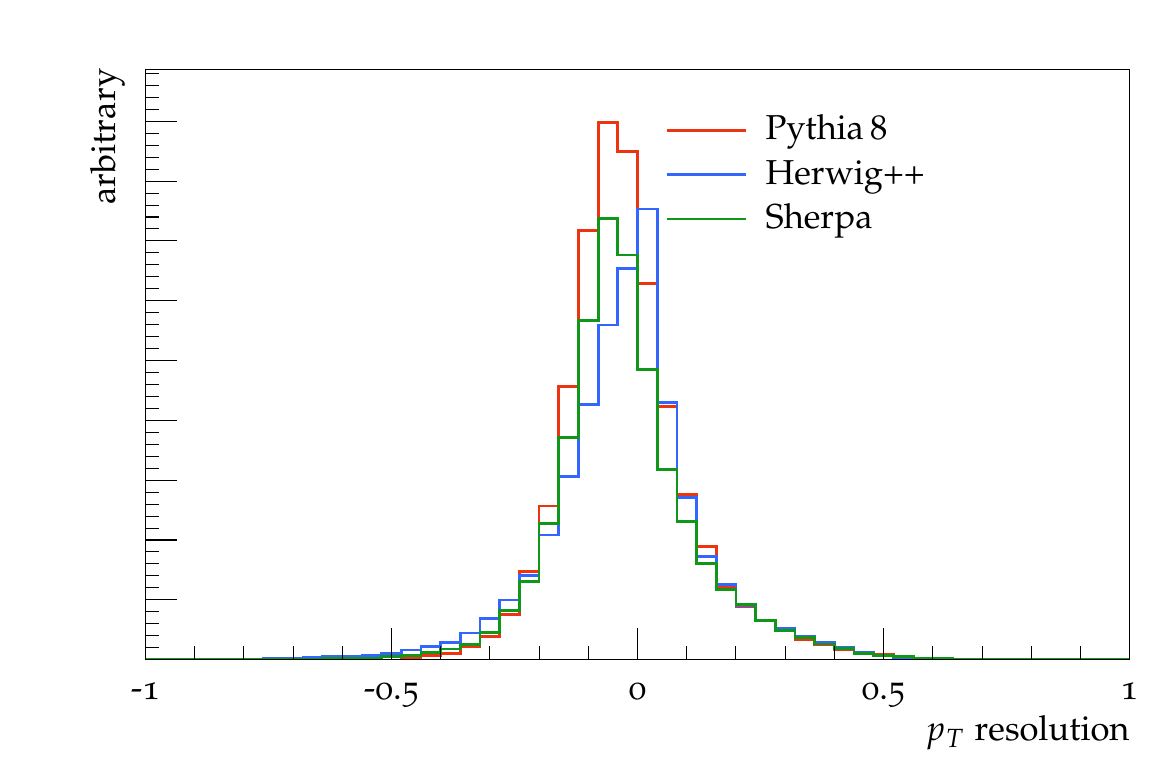}\\
  \img[0.48]{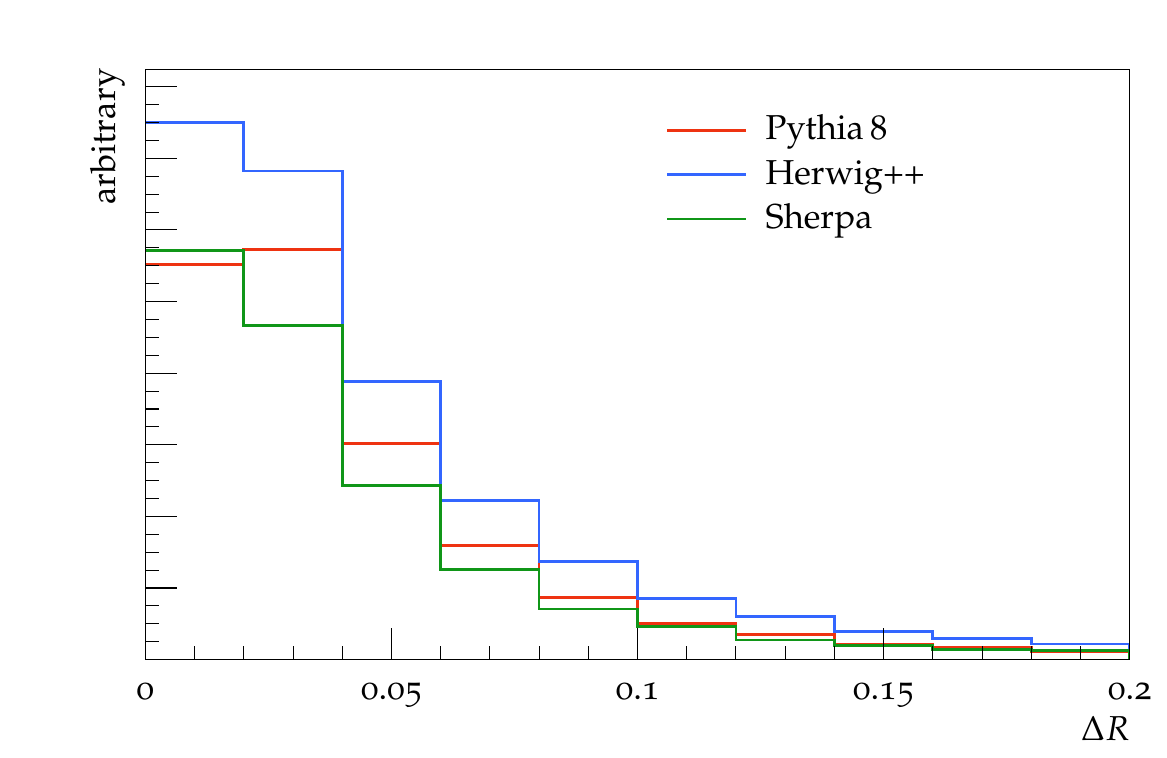} \quad \img[0.48]{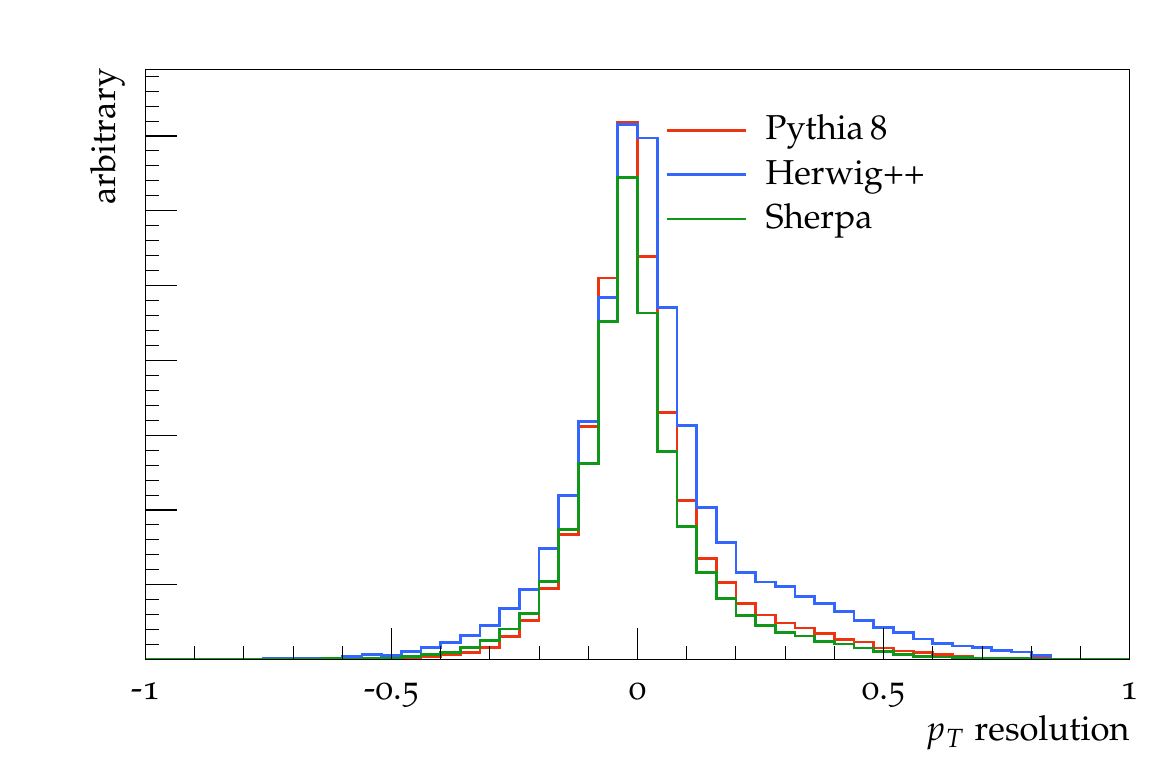}
  \caption{Light tagging performance/comparisons: inclusive jet events, with
    gluon-labelled jets on the top row and light-quark-labelled jets on the
    bottom row. Particle jets clustered using the \akt algorithm, labelling
    parton jets with the \kt algorithm.}
  \label{fig:jets2cmplight}
\end{figure}
\begin{figure}[p]
  \centering
  \img[0.48]{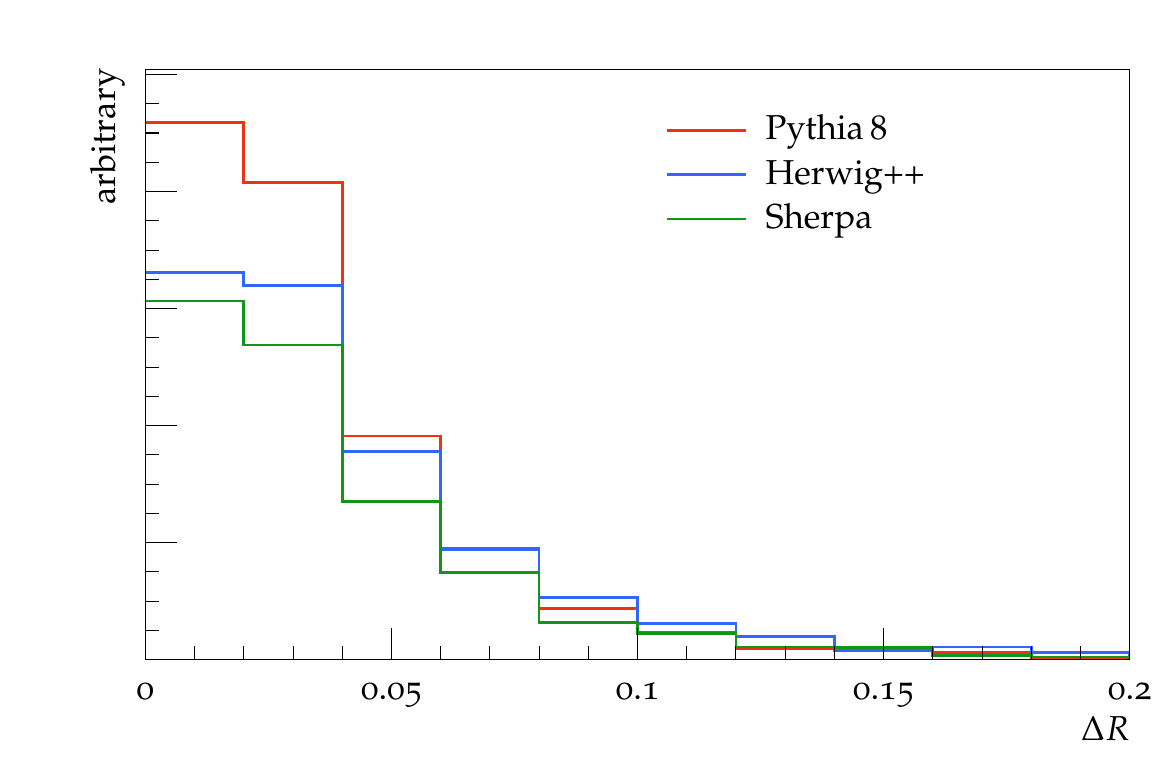} \quad \img[0.48]{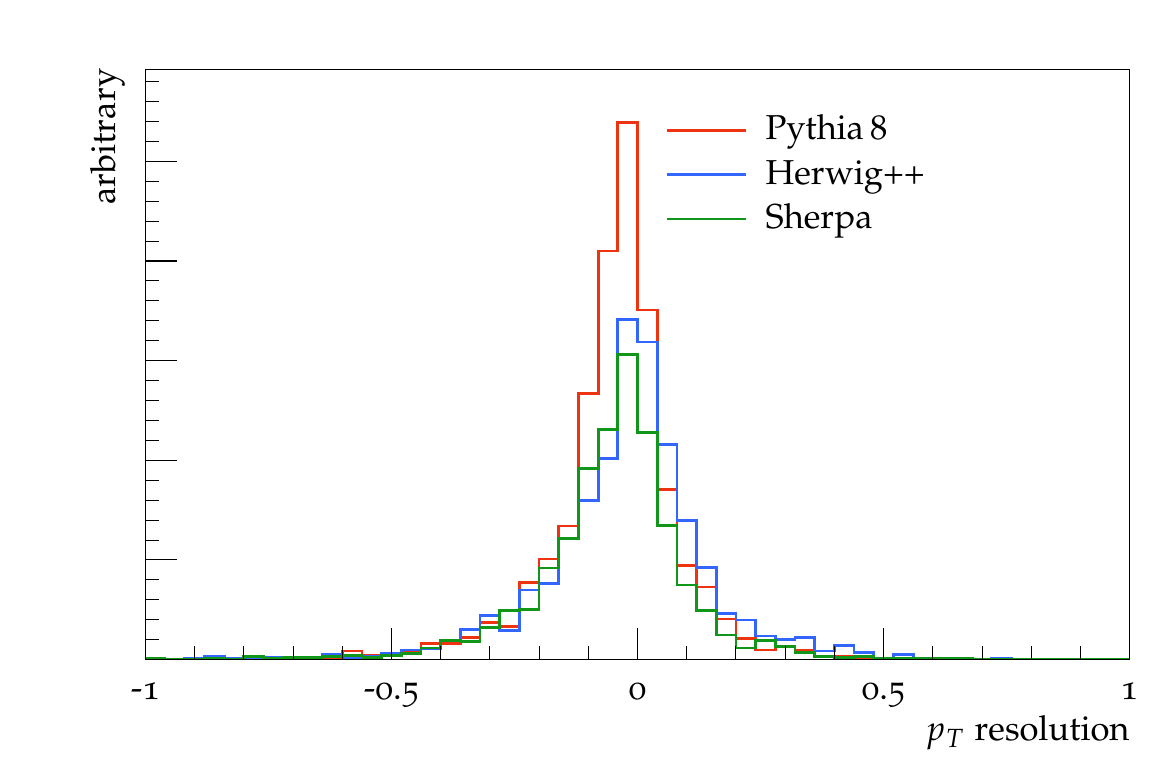}\\
  \img[0.48]{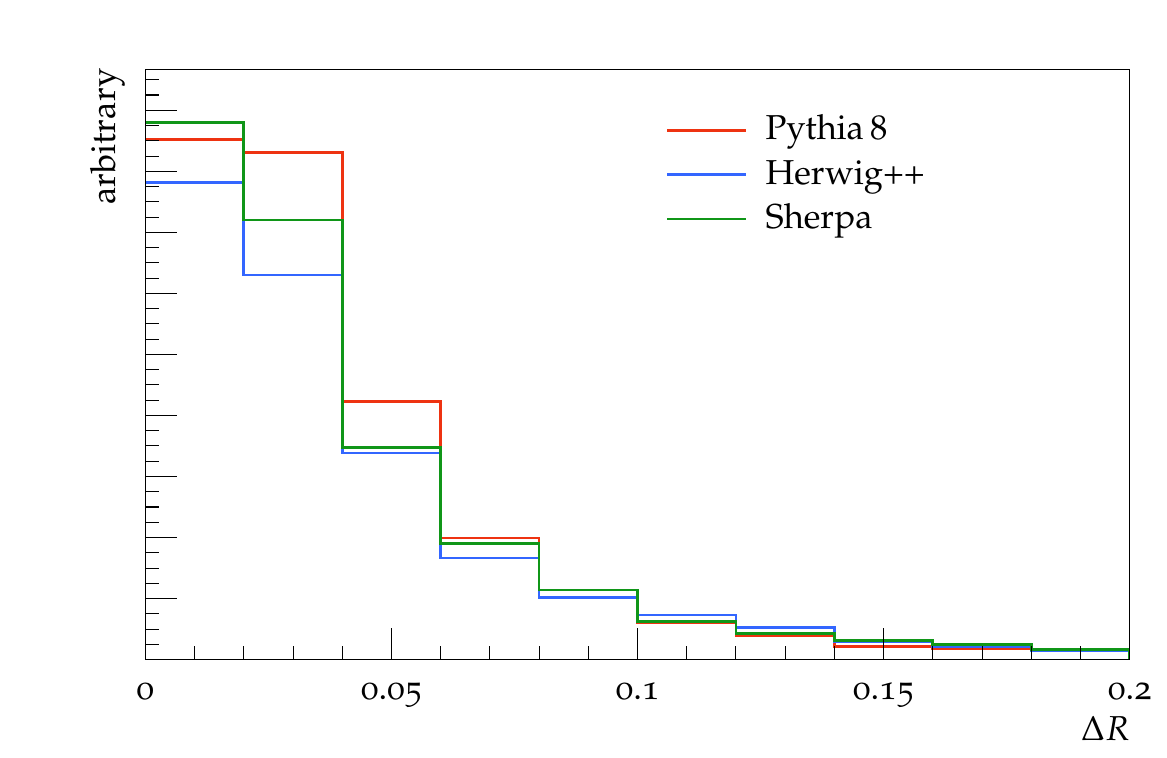} \quad \img[0.48]{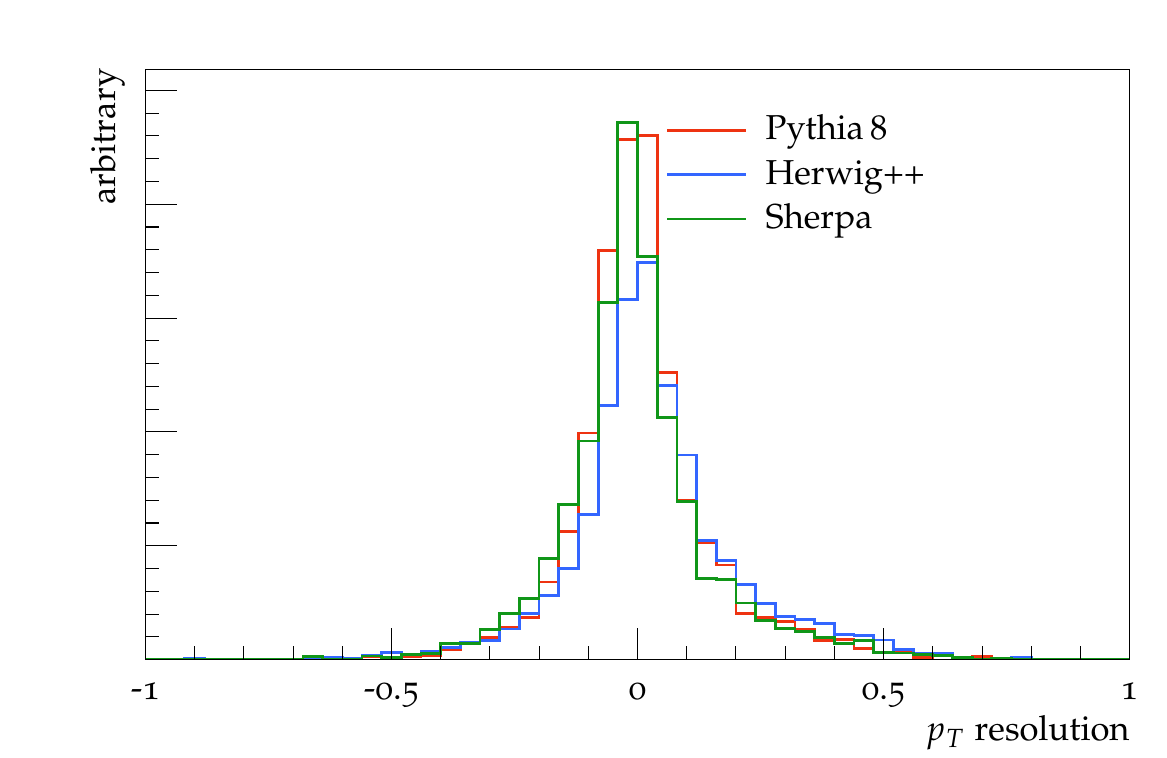}
  \caption{Heavy tagging performance/comparisons: inclusive jet events, with
    bottom-labelled jets on the top row and charm-labelled jets on the bottom
    row. Particle jets clustered using the \akt algorithm, labelling parton jets
    with the \kt algorithm.}
 \label{fig:jets2cmpheavy}
\end{figure}
\begin{figure}[p]
  \centering
  \img[0.48]{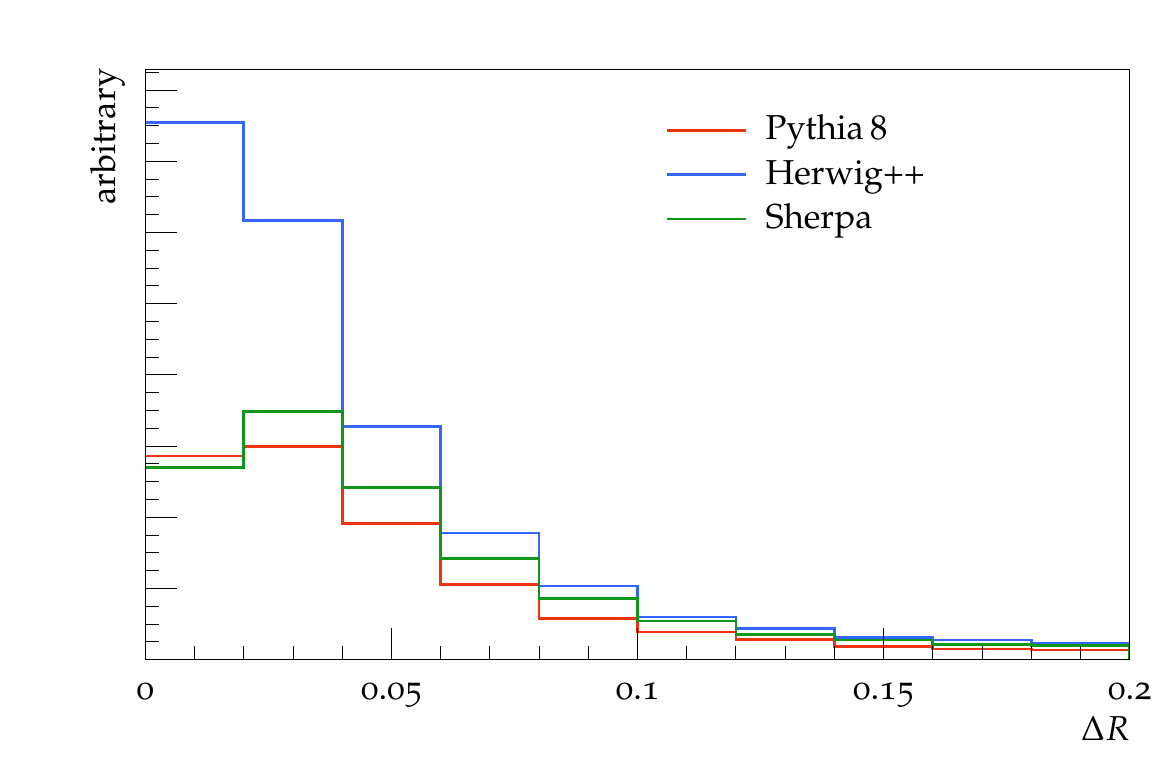} \quad \img[0.48]{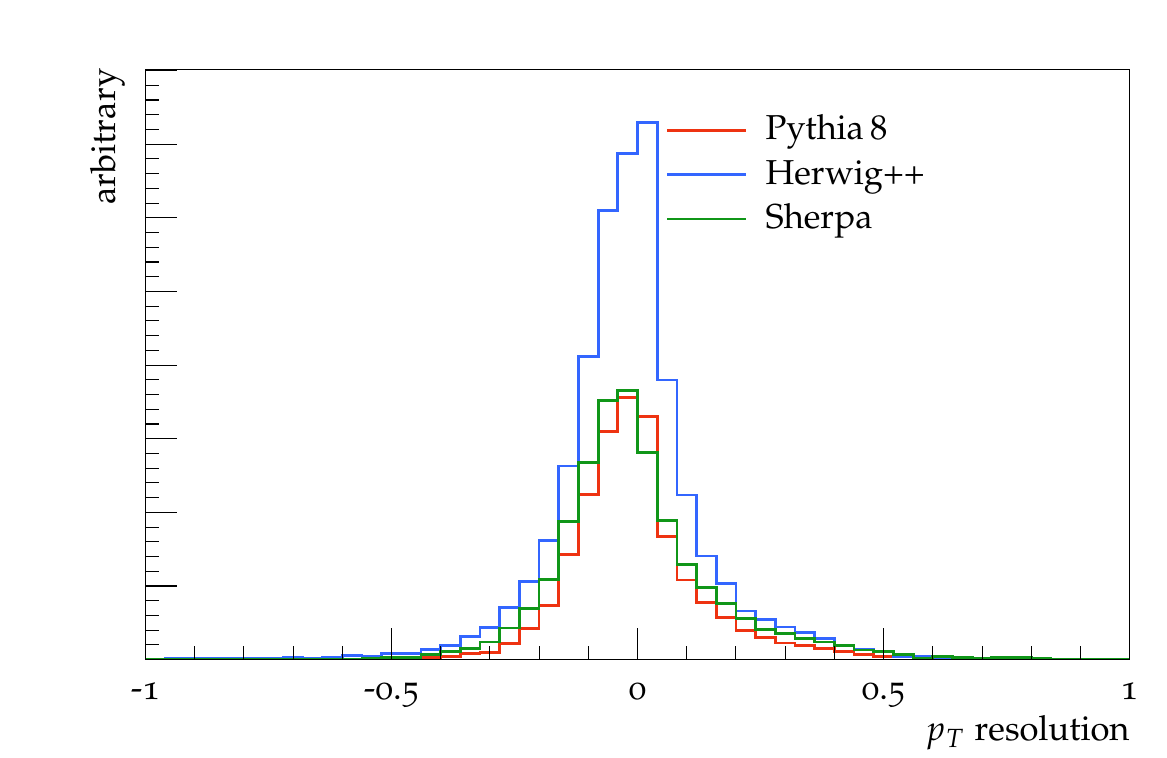}\\
  \img[0.48]{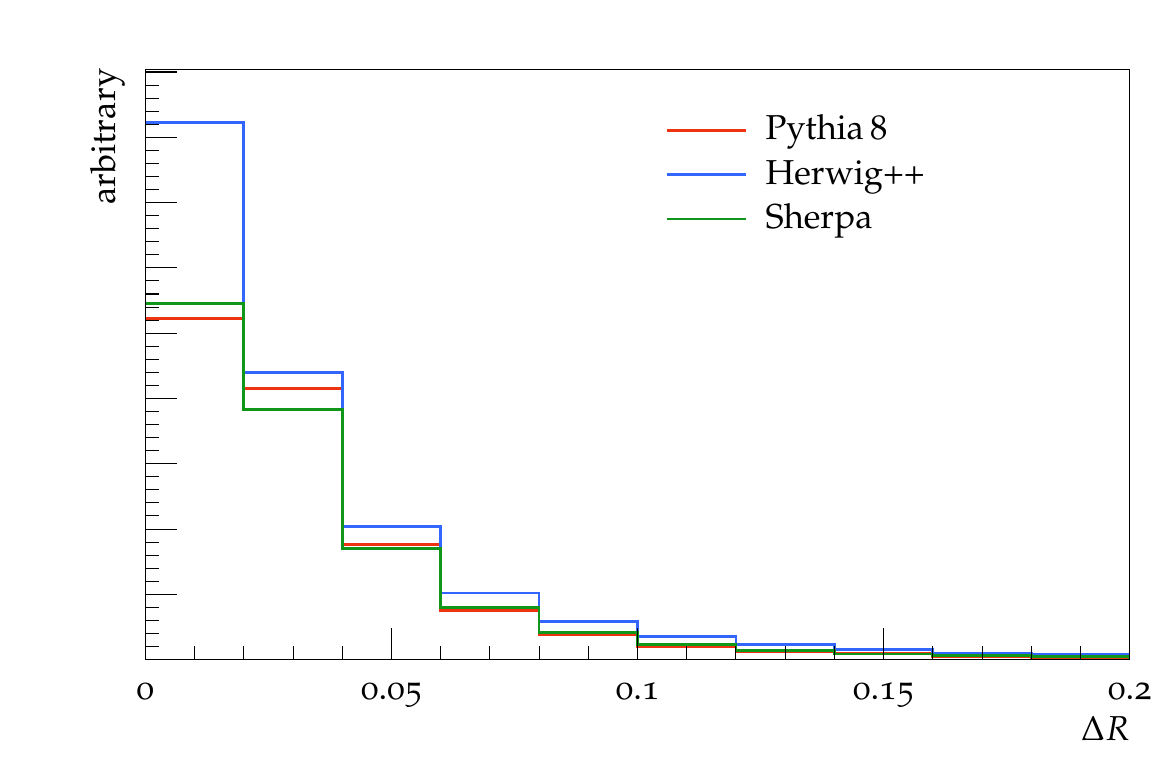} \quad \img[0.48]{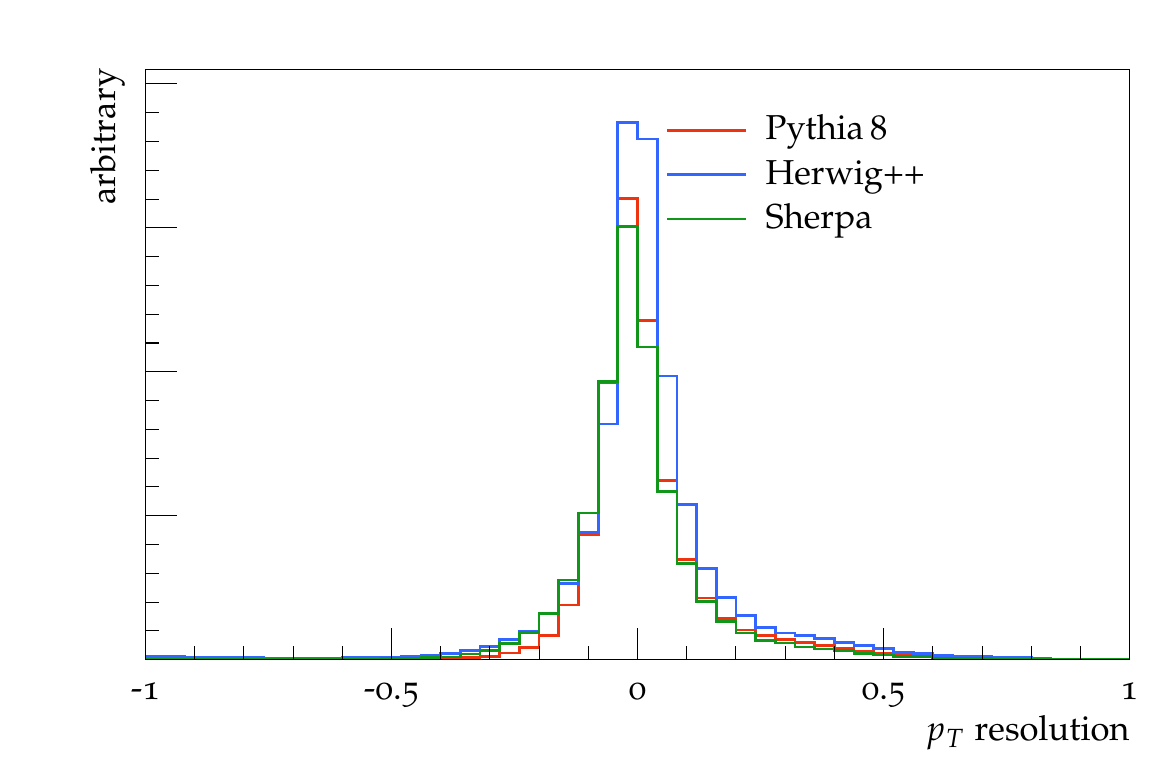}\\
  \img[0.48]{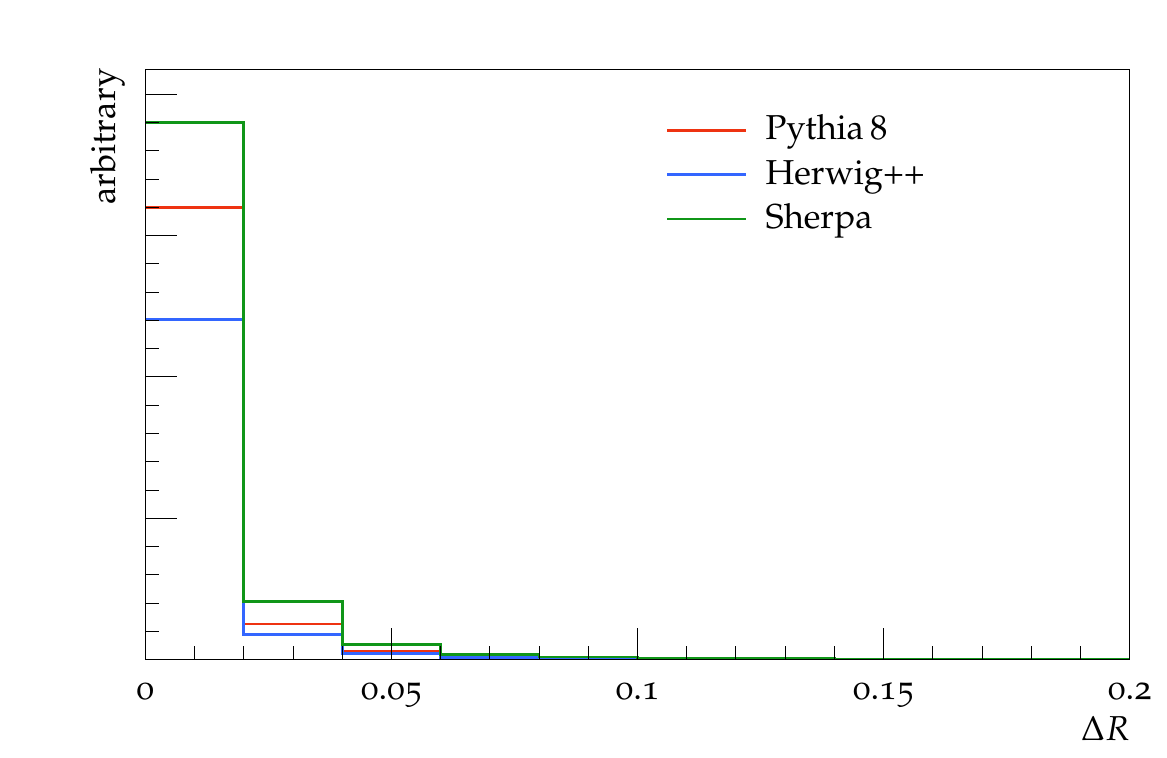} \quad \img[0.48]{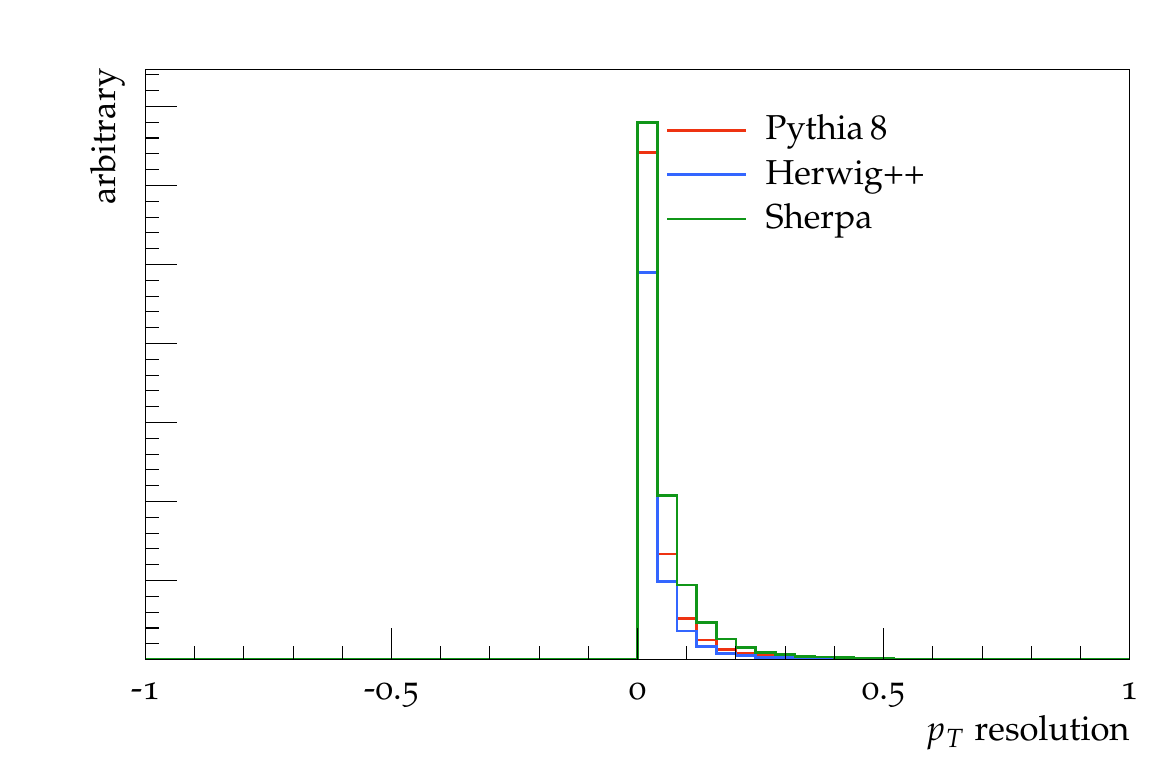}
  \caption{Light tagging performance/comparisons: $\gamma + \text{jet}$ events,
    with gluon-labelled jets on the top row, light-quark-labelled jets in the
    middle row, and isolated prompt photons on the bottom row. Particle jets
    clustered using the \akt algorithm, labelling parton jets with the \kt
    algorithm. The one-sided \pt resolution distribution for the photon-labelled
    jet is by construction as described in the text.}
 \label{fig:gammajetcmplight}
\end{figure}

In Figures~\ref{fig:jets2cmplight} and~\ref{fig:jets2cmpheavy} we show the
\pt~resolution, $\Delta\pt/\pt = \left(\pt^\text{jet} -
  \pt^\text{label}\right)/\pt^\text{jet}$ and $\Delta{R}$ measure between the
particle jets and their assigned labels in the inclusive-QCD-jet process. The
light parton measures are shown for the photon+jet hard process type in
Figure~\ref{fig:gammajetcmplight}, the less interesting heavy label performance
being relegated to Appendix~\ref{app:extraplots}. A \kt measure has been used in the construction of
the QCD-aware labelling jets, but the particle jets are \akt as is standard for
the LHC experiments; the effect of this mismatch will be investigated in
\Fref{sec:measurecmp}. 
%
The effect of ME scale dependence on the plot normalisation has been largely
eliminated by rescaling all histograms to correspond to the Pythia\,8
cross-section calculation. 

The main point of note in these plots is general consistency between generators,
and good-quality kinematic matching of the labels to the particle jets. The
consistency is not perfect -- for example Pythia events feature significantly
more bottom-labelled jets than either of the other generators in the jets
process, and Herwig\xx produces many more gluon-labelled jets in the
$\gamma+\text{jet}$ event type -- but otherwise these resolutions are very
compatible both with each other and with the expectations of approximate
parton--jet duality. Where shape differences are seen, e.g. the Herwig\xx high
tail for light-quark-labelled $\Delta\pt/\pt$ in jet events, they are
significantly suppressed relative to the peak of the distribution; addition of a
\pt resolution window cut to the labelling algorithm could remove the few
anomalous labels assigned from long tails such as this.

The one-sided $\Delta\pt$ distribution for the $\gamma+\text{jet}$ samples is
deserving of explanation.  This feature is by construction: to be labelled as a
photon, the labelling ``parton'' is the exact same final-state photon as will
enter the particle-jet finding, because there is no parton that a photon can
cluster with, without losing its flavour. Hence the total jet \pt must be at
least the same as the labelling photon: momentum can only be added, not
subtracted, within the jet cone. By contrast, for quark or gluon jets there are
non-perturbative processes such as hadronisation and colour-reconnection which
can reduce the particle-jet energy below that of their labelling parton. It is
likely that an asymmetry of similar size is convolved into the $\Delta\pt/\pt$
distributions for quark and gluon jets, but that their overall greater width
from non-perturbative modelling dominates the roughly symmetric peak shape.


\subsubsection{Flavour label ratios}
\label{sec:ratios}





\begin{table}[tbp]
  \centering
  \begin{tabular}{ll@{\qquad}r@{\qquad}rr}
    \toprule
            &           & Jets  & \multicolumn{2}{c}{$\gamma + \text{jet}$} \\
    Scheme  & Generator & $q/g$ & $\gamma/g$ & $q/g$ \\
    \midrule
    Max-\pt & Pythia\,8 & 0.38  & 17.2 & 10.5 \\
            & Herwig\xx  & 0.33  &  7.7 &  4.8 \\
            & Sherpa    & 0.55  & 21.0 &  9.6 \\
    \addlinespace
    \kt     & Pythia\,8 & 0.80  & 10.4 &  8.2 \\
            & Herwig\xx  & 1.17  &  3.6 &  4.6 \\
            & Sherpa    & 0.85  & 10.5 &  7.5 \\
    \addlinespace
    \akt    & Pythia\,8 & 0.79  & 10.2 &  8.3 \\
            & Herwig\xx  & 1.74  &  3.2 &  4.5 \\
            & Sherpa    & 0.86  & 10.2 &  7.5 \\
    \addlinespace
Reclustered & Pythia\,8 & 0.77  & 10.1 &  8.0 \\
            & Herwig\xx  & 1.36  &  3.5 &  4.8 \\
            & Sherpa    & 0.83  & 10.1 &  7.3 \\
    \bottomrule
  \end{tabular}
  \caption{Jet label ratios for the combined sample of leading and subleading jets
    constructed in inclusive jet and $\gamma+\text{jet}$ simulated events, for
    various MC generators.}
  \label{tab:ratios}
\end{table}





The other feature of these plots immediately worth commenting on is how the
total number of jets labelled as gluon or quark in these two samples compare to
the cross-sections of the fixed-order subprocess matrix elements.  We compute
these ratios from the integral of each distribution for the two leading jets
only, to ensure final-state quantities comparable to the 2-particle matrix
element final states, and are not biased by the propensities of the generators
to produce different numbers of above-threshold jets. The summary ratios are
presented in \Fref{tab:ratios}, and we now discuss them for the two process
types.

\begin{description}
\item[Inclusive jet sample:] the Pythia\,8 leading-order matrix elements had the
  following cross-sections: $\sigma(gg \to gg) = 2.71\;\mb$, $\sigma(qg \to qg)
  = 1.41\;\mb$, $\sigma(qq \to qq) = 0.15\;\mb$, $\sigma(gg \to q\bar{q}) =
  0.07\;\mb$. Summing these according to their contributions to the rates of
  final state quark or gluon partons, a fixed-order quark/gluon jet ratio of
  0.27 is expected.

  Comparing the normalisations of the simulated gluon and quark observables
  gives QCD-aware \kt quark/gluon labelling ratios of between 0.8--1.2 for the three
  shower generators, with Herwig\xx at 1.17 a clear outlier from the other two
  generators near 0.8. The equivalent range for the max-\pt labelling scheme is
  0.33--0.55. There is hence a substantial different between the $q/g$ ratios
  obtained from max-\pt and QCD-aware schemes, the max-\pt scheme giving
  label rates closer to the fixed-order expectation than QCD-aware -- although this
  should not be overinterpreted as indicting (in)correctness of either scheme.

  The inverted $q/g$ ratio from Herwig\xx is due to an excess of quarks, clearly
  visible in the light-quark plots of \Fref{fig:jets2cmplight}. Requiring a
  jet-label \pt match by cuts on the tails of \pt resolution brings the Herwig\xx
  $q/g$ ratio to 1.0 -- closer to the others, but still not good agreement.

\item[$\gamma+\text{jet}$ sample:] applying the same methodology as for the
  inclusive jet ratios, the two tree-level subprocess cross-sections are
  $\sigma(qg \to q\gamma) = 650\;\nb$ and $\sigma(q\bar{q} \to g\gamma) =
  53\;\nb$, corresponding to expected fixed-order photon/gluon and quark/gluon
  ratios of 13.2 and 12.2.

  The fully showered QCD-aware \kt ratios are between 3.6--10.4 and 4.6--8.2
  respectively, and the max-\pt equivalents are 7.7--21.0 and 4.8--10.5. In both
  schemes, there is agreement between the rates for Pythia8 and Sherpa which are
  at the high ends of the ratio ranges for both processes, but Herwig\xx is
  highly discrepant. The much lower $\gamma/g$ and $q/g$ rates for Herwig\xx
  appear to be driven by its very high gluon rate, clearly visible in
  \Fref{fig:gammajetcmplight}; there is this time no obvious matching cut on
  $\Delta{R}$ or \pt resolution which could address this issue.

  %
\end{description}

We stress again that these predictions are subject to significant resummation
corrections and hence do not identify the ``right answer''. The broad
expectations from fixed order subprocess cross-sections are met by Sherpa and
Pythia8, but Herwig\xx significantly deviates from the other generators by producing
unusually many quark jets in jet events, and gluon jets in $\gamma+\text{jet}$ events.
We will return to these features later.


\subsection{Dependence on clustering distance measure}
\label{sec:measurecmp}


In our introduction of the QCD-aware method, we motivated partonic clusterings
by analogy to a ``rewinding'' of QCD evolution through gluon radiation (and the
analogous evolution for photon emission).  This was also the historical
motivation for the flavour-blind ``bland'' \kt clustering, but it has since
proven useful to use alternative measures, most notably the \akt distance which
has no clear link to resummation.  We here entertain the possibility that the
\akt distance measure may prove to have useful properties despite its relative
lack of \emph{a priori} motivation.

To reduce the potential for kinematic mismatch of \kt and \akt jet shapes, we
also consider a \kt-based QCD-aware reclustering of final partons matched to
\akt particle jets.  In this approach, final partons are first
ghost-associated~\cite{Cacciari:2007fd} as part of the construction of \akt
particle jets, then for each particle jet the collection of associated final
partons is clustered using the QCD-aware \kt measure (``\kt-reclustered'').  The
\akt particle jet then inherits the label of the \kt final parton jet with the
smallest \dr separation to its axis.

\begin{figure}[tp]
  \centering
  \img[0.48]{figs/jets2/Inclusive_Gluon_Kt_Dpt.pdf} \quad \img[0.48]{figs/jets2/Inclusive_Light_Kt_Dpt.pdf}\\
  \img[0.48]{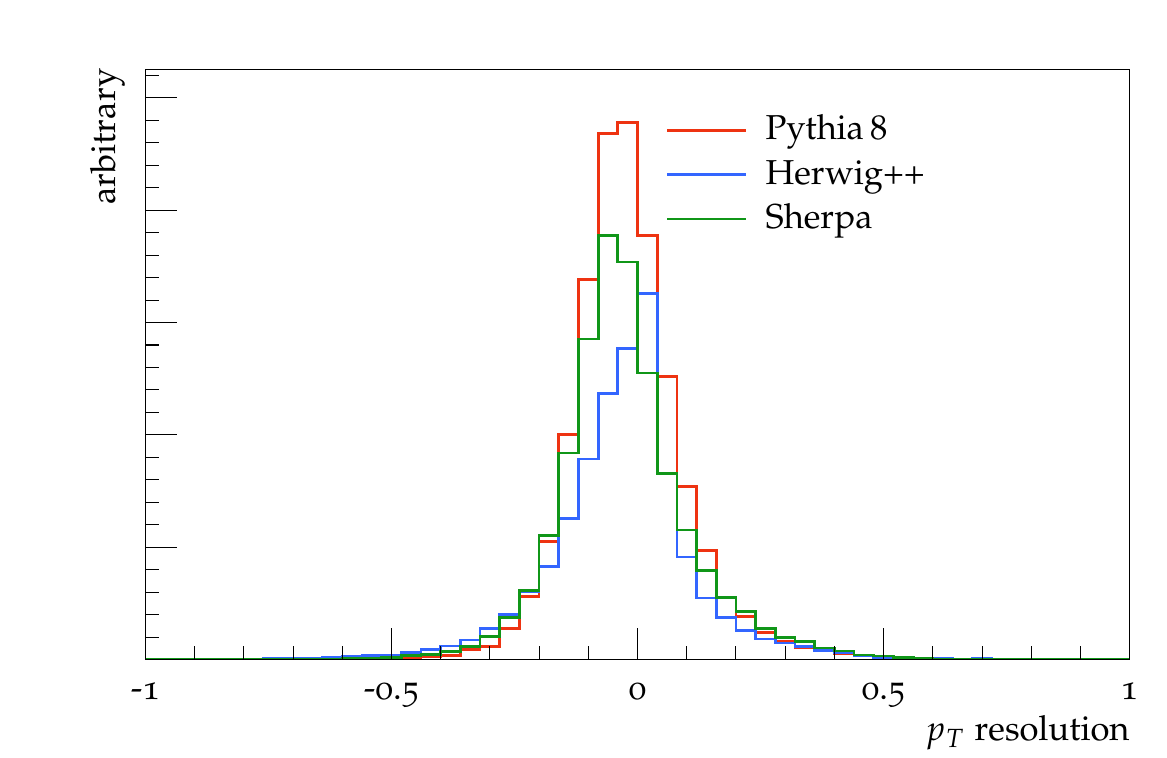} \quad \img[0.48]{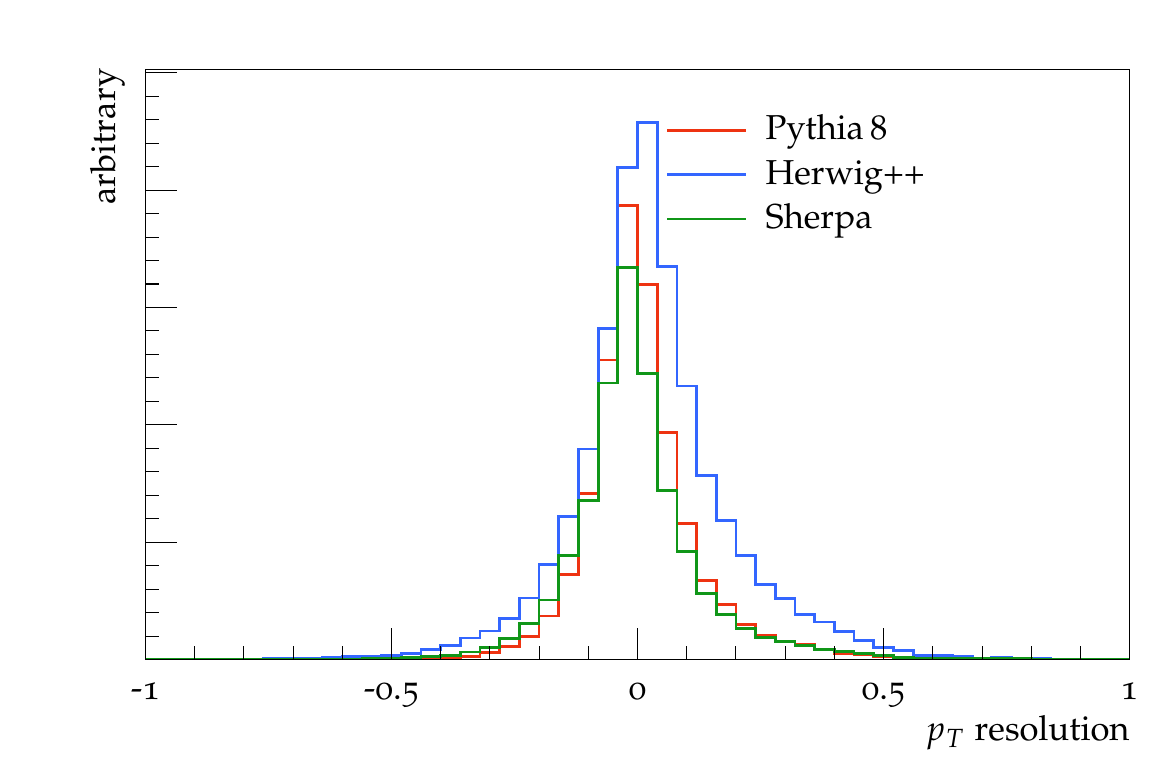}\\
  \img[0.48]{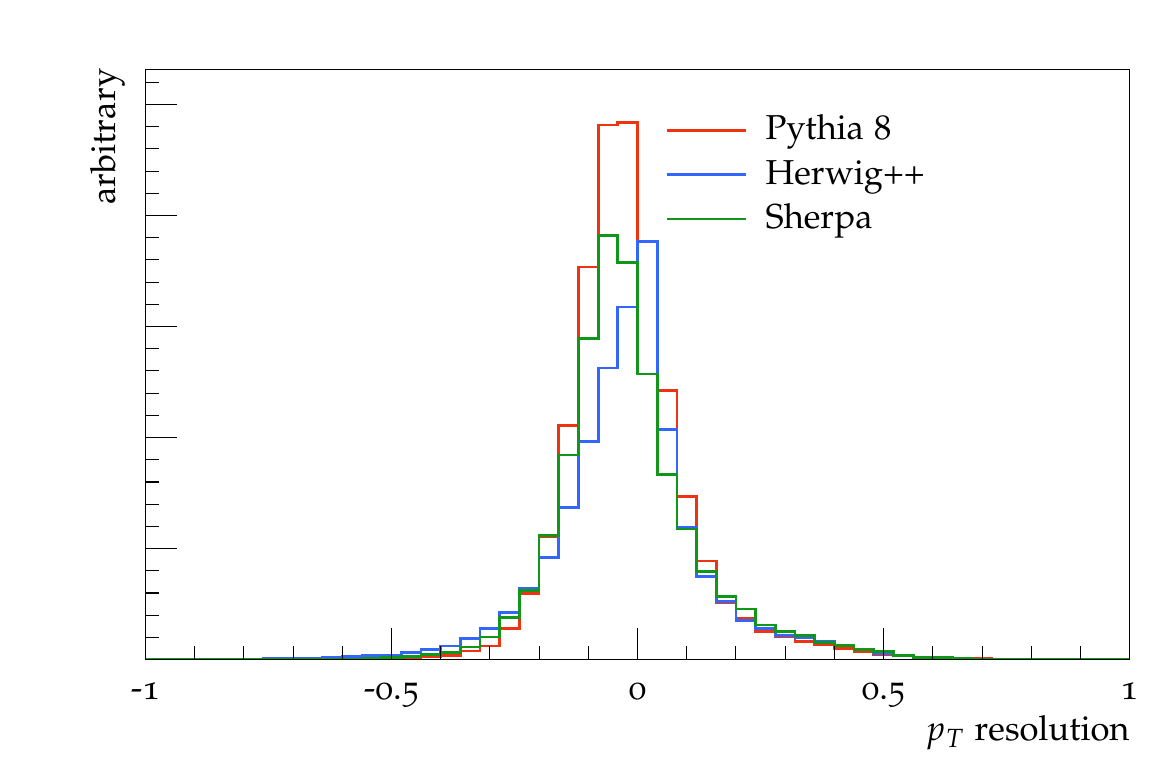} \quad \img[0.48]{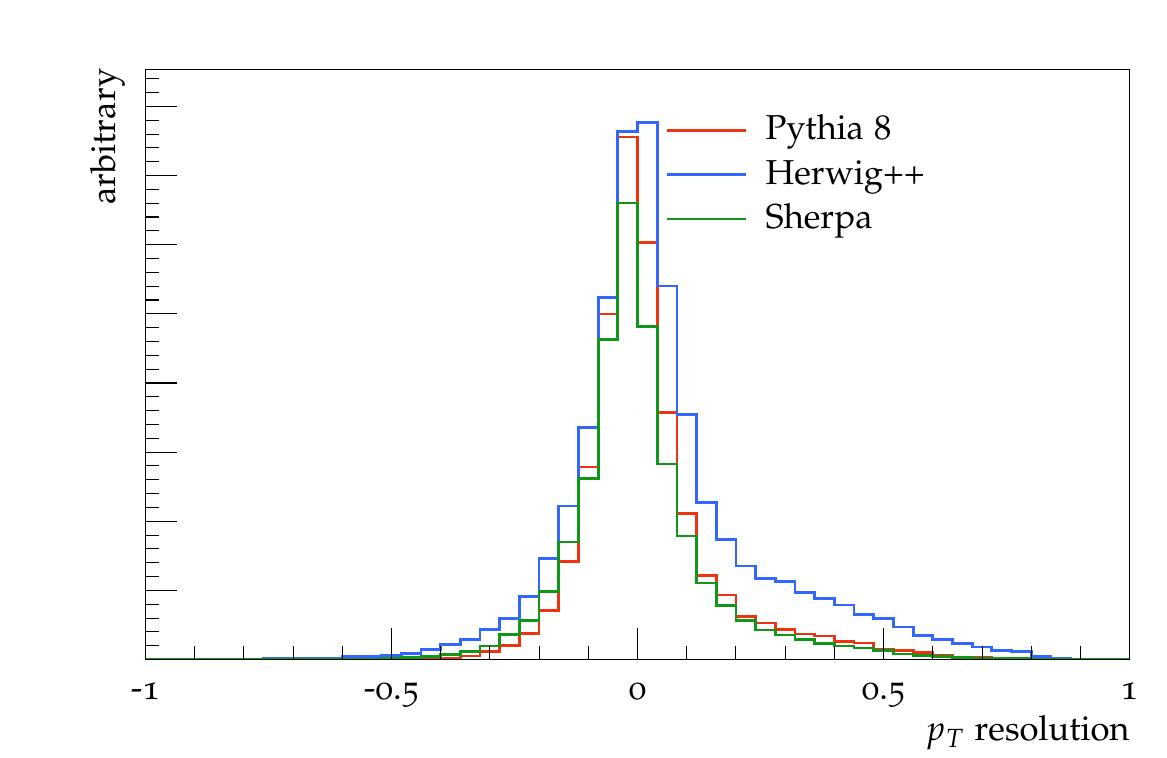}\\
  \caption{$\Delta\pt/\pt$ performance comparisons for gluon- (left)
      and light quark-labelled jets (right) in inclusive jet events,
      showing \kt labeled jets on the top row, anti-\kt in the middle,
      and \kt-reclustered on the bottom.}
  \label{fig:labelcmpdpt}
\end{figure}

\begin{figure}[tp]
  \centering
  \img[0.48]{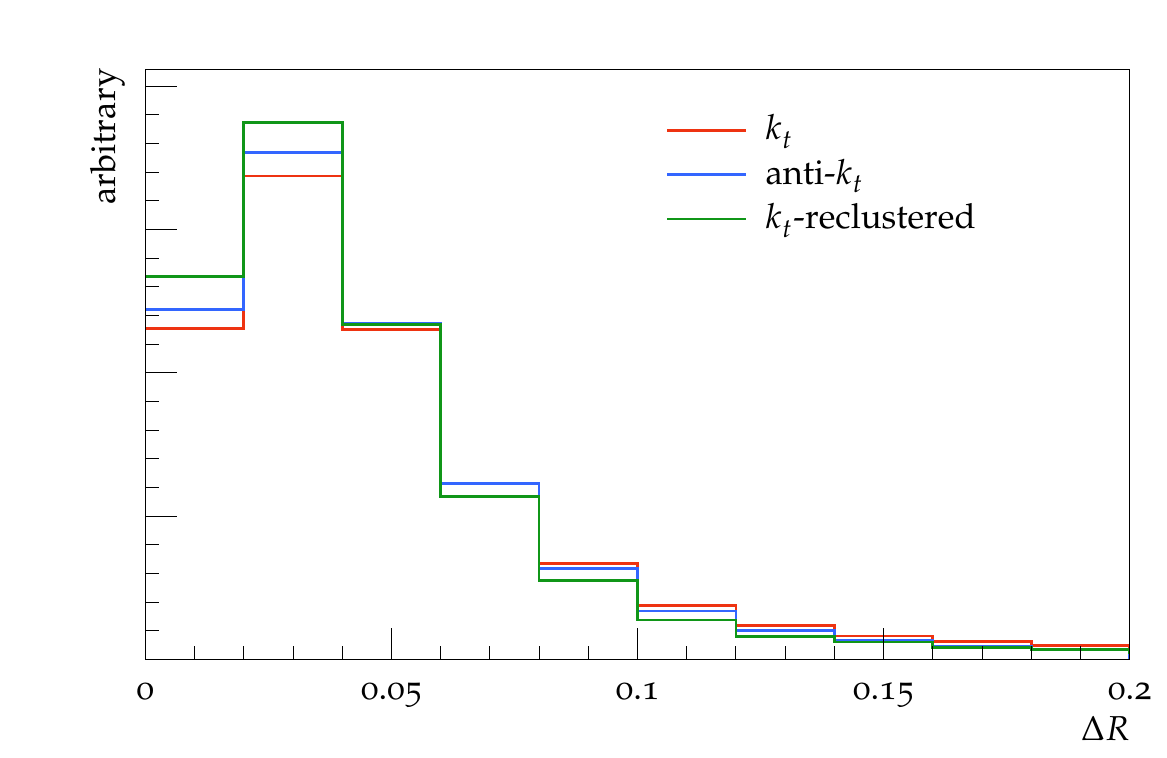} \quad
  \img[0.48]{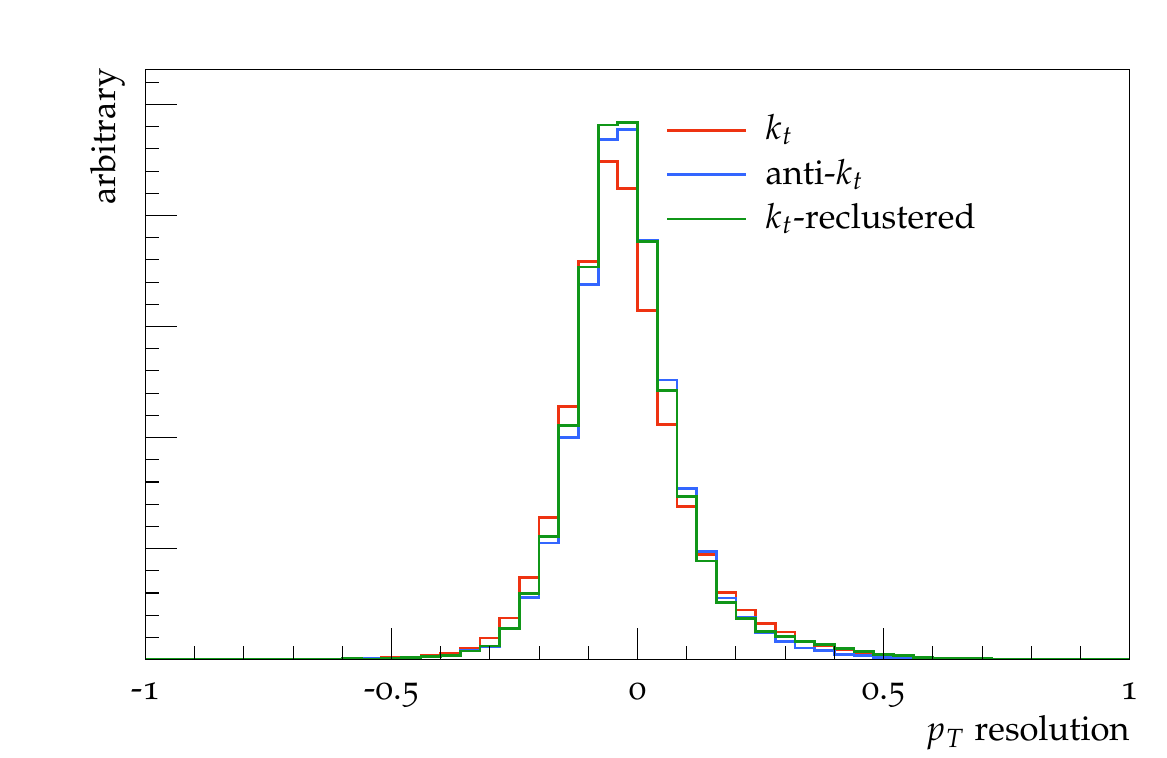}\\
  \img[0.48]{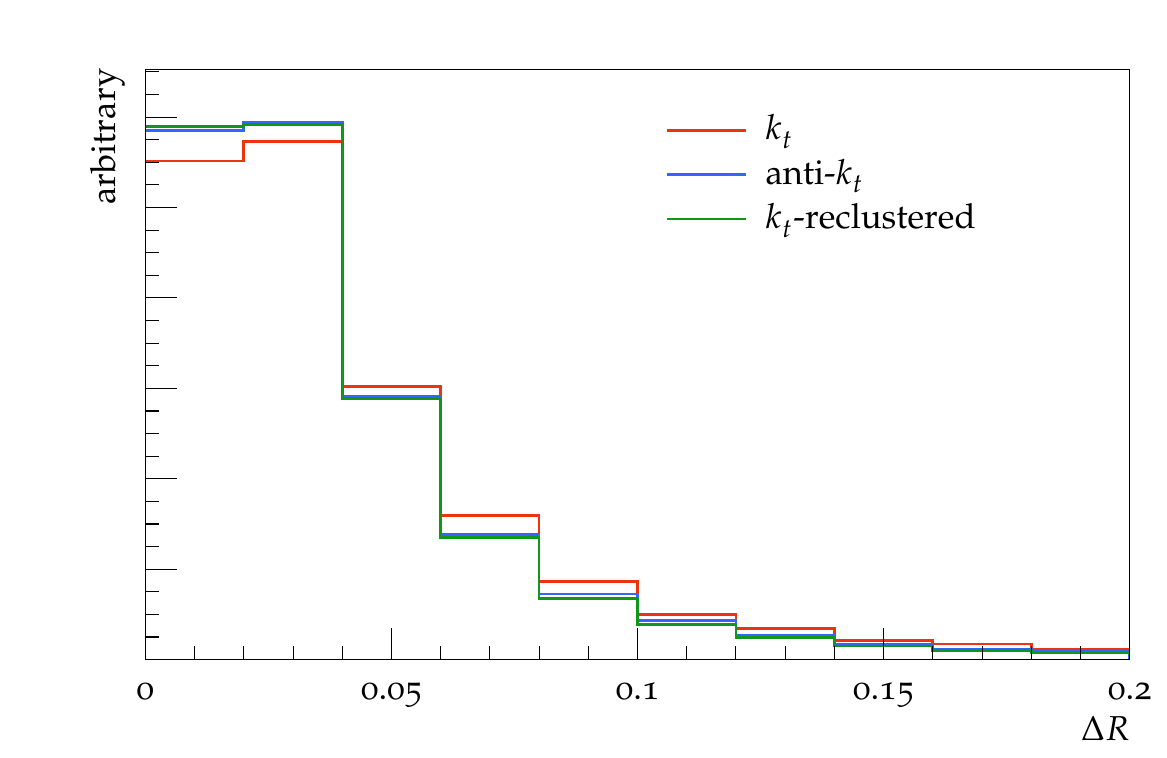} \quad
  \img[0.48]{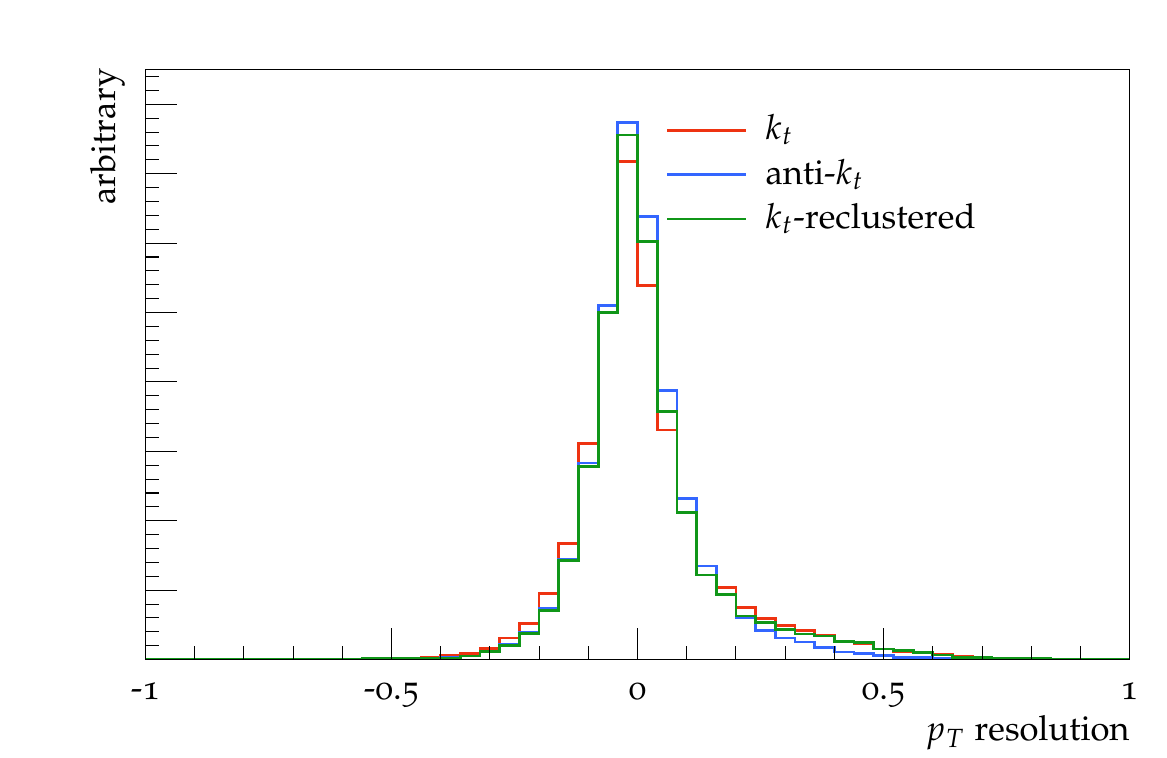}\\
  \caption{$\Delta{R}$ (left) and $\Delta\pt/\pt$ (right) clustering
      performance comparisons for gluon (top) and light (bottom)
      labelled jets in Pythia inclusive jet events.}
  \label{fig:pythialabelcmp}
\end{figure}

\begin{table}[t]
  \centering

  {\small
  \begin{tabular}{cr|rrrrrr}
    \toprule
    \multicolumn{2}{c}{ } & \multicolumn{6}{c}{$k_t$ label} \\
    \multicolumn{2}{c|}{ } & none & $g$ & $q$ & $c$ & $b$ & $\gamma$ \\
    \cline{2-8}
    \multirow{6}{*}{\rotatebox{90}{anti-$k_t$ label}} & none & 0.8 & 0.1 & - & - & - & - \\
     & $g$ & 0.3 & 52.5 & 2.5 & 0.1 & - & - \\
     & $q$ & 0.1 & 2.1 & 33.7 & - & - & - \\
     & $c$ & - & - & - & 5.0 & - & - \\
     & $b$ & - & - & - & - & 2.5 & - \\
     & $\gamma$ & - & - & - & - & - & - \\
     \bottomrule
  \end{tabular}
  \hfill
  \begin{tabular}{cr|rrrrrr}
    \toprule
    \multicolumn{2}{c}{ } & \multicolumn{6}{c}{$k_t$-reclustered label} \\
    \multicolumn{2}{c|}{ } & none & $g$ & $q$ & $c$ & $b$ & $\gamma$ \\
    \cline{2-8}
    \multirow{6}{*}{\rotatebox{90}{$k_t$ label}} & none & - & 0.9 & 0.3 & - & - & - \\
     & $g$ & - & 53.0 & 1.6 & - & - & - \\
     & $q$ & - & 2.5 & 33.8 & - & - & - \\
     & $c$ & - & 0.1 & - & 5.0 & - & - \\
     & $b$ & - & - & - & - & 2.5 & - \\
     & $\gamma$ & - & - & - & - & - & - \\
     \bottomrule
  \end{tabular}

  \vspace{2em}

  \begin{tabular}{cr|rrrrrr}
    \toprule
    \multicolumn{2}{c}{ } & \multicolumn{6}{c}{max-$p_T$ label} \\
    \multicolumn{2}{c|}{ } & none & $g$ & $q$ & $c$ & $b$ & $\gamma$ \\
    \cline{2-8}
    \multirow{6}{*}{\rotatebox{90}{$k_t$ label}} & none & - & 1.1 & 0.1 & - & - & - \\
     & $g$ & - & 53.5 & 1.1 & - & - & - \\
     & $q$ & - & 15.7 & 20.5 & - & - & - \\
     & $c$ & - & 1.5 & - & 3.6 & - & - \\
     & $b$ & - & 0.4 & - & - & 2.1 & - \\
     & $\gamma$ & - & - & - & - & - & - \\
     \bottomrule
  \end{tabular}
  } 

  \caption{Correlation matrices for pairs of labelling schemes in
      Pythia inclusive jet events.
      Each entry shows the fraction of all jets (in percent) given
      a pair of labels by the schemes listed on the vertical and
      horizontal axes.
      $q$ here denotes a light-quark label.
      Fractions less than 0.1\% are replaced with a dash.
  }
  \label{tab:pythialabelcorrelationmatrices}
\end{table}

\begin{table}[t]
  \centering

  {\small
  \begin{tabular}{cr|rrrrrr}
    \toprule
    \multicolumn{2}{c}{ } & \multicolumn{6}{c}{max-$p_T$ label} \\
    \multicolumn{2}{c|}{ } & none & $g$ & $q$ & $c$ & $b$ & $\gamma$ \\
    \cline{2-8}
    \multirow{6}{*}{\rotatebox{90}{$k_t$ label}} & none & - & 0.2 & 0.3 & - & - & 0.6 \\
     & $g$ & - & 5.6 & 2.7 & 0.2 & - & 2.8 \\
     & $q$ & - & 4.3 & 25.6 & 0.2 & - & 14.7 \\
     & $c$ & - & 0.2 & - & 6.1 & - & 0.3 \\
     & $b$ & - & - & - & - & 0.9 & - \\
     & $\gamma$ & - & - & - & - & - & 34.8 \\
     \bottomrule
  \end{tabular}
  \hfill
  \begin{tabular}{cr|rrrrrr}
    \toprule
    \multicolumn{2}{c}{ } & \multicolumn{6}{c}{max-$p_T$ label} \\
    \multicolumn{2}{c|}{ } & none & $g$ & $q$ & $c$ & $b$ & $\gamma$ \\
    \cline{2-8}
    \multirow{6}{*}{\rotatebox{90}{$k_t$ label}} & none & - & - & 0.2 & - & - & 0.9 \\
     & $g$ & - & 3.1 & 1.8 & 0.2 & - & 0.4 \\
     & $q$ & - & 0.9 & 27.5 & - & - & 3.4 \\
     & $c$ & - & - & - & 6.4 & - & 0.2 \\
     & $b$ & - & - & - & - & 0.9 & - \\
     & $\gamma$ & - & - & - & - & - & 53.8 \\
     \bottomrule
  \end{tabular}
  } 

  \caption{Correlation matrices for pairs of labelling schemes in
      Herwig\protect\xx $\gamma+\text{jet}$ events with MPI on (left) and MPI off (right).
      Each entry shows the fraction of all jets (in percent) given
      a pair of labels by the schemes listed on the vertical and
      horizontal axes.
      $q$ here denotes a light-quark label.
      Fractions less than 0.1\% are replaced with a dash.
    }
  \label{tab:herwigmpilabelcorrelationmatrices}
\end{table}

In Figure~\ref{fig:labelcmpdpt} the $\Delta{\pt}/\pt$ performance
measure is again shown for the three shower
MC generators, for gluon- and quark-labelled jets, respectively.
The $\Delta{R}$ performance comparisons are shown in
Figure~\ref{fig:labelcmpdr} in the appendix.
Figure~\ref{fig:pythialabelcmp} shows a direct comparison of these
spectra across labelling schemes for Pythia inclusive jet events.
Perhaps surprisingly, the differences in shape are small, even though
\akt is expected to have a virtually meaningless cluster sequence.
There are two explanations for this: first, the flavour combination
rules mean that there is much less difference in the clustering
sequence than would be the case in a normal flavour-blind \kt vs. \akt
clustering comparison; and second, the resulting labels often depend
not on the detailed order of clusterings but on \emph{whether} an
unpaired (anti)quark ever gets clustered into an otherwise
gluon-flavoured parton pseudojet.
Clustering of gluons or photons on to a quark
or lepton-flavoured pseudojet changes only kinematics, not the label, and the
change to the kinematics is invariant of the order of the boson clustering --
the effect of measure is hence largely relegated to clusterings near the jet
radius and those which occur during the differing periods during which the
pseudojet kinematics are stabilising.

How the labels of individual particle jets differ between
schemes is also of interest. For instance, the \kt scheme may
assign a jet a gluon label while the \akt scheme may label it as a
quark. This is summarised in the label correlation matrices in
Table~\ref{tab:pythialabelcorrelationmatrices}. 
The anti-\kt and \kt-reclustered schemes are in
the top row, while the bottom-left matrix compares QCD-aware \kt with
the max-\pt scheme, in which a particle jet inherits the label of the
highest-\pt ghost-associated parton from \emph{any} step of the QCD shower
evolution. Labelling algorithms similar to the max-\pt scheme have
been used in performance studies at ATLAS~\cite{Aad:2011he}.

Table~\ref{tab:herwigmpilabelcorrelationmatrices} gives some
indication of the origin of the difference between max-\pt and QCD-aware
labelling for Herwig\xx $\gamma+\text{jet}$ events: there is a very significant
off-diagonal contribution in the $q$--$\gamma$ cell, i.e. jets which are
identified as photons in the max-\pt scheme, but as quarks in the QCD-aware
scheme. This implies that low-momentum quarks in the vicinity of the hard photon
are ``capturing'' that hard object and stealing its photon label. The unanswered
question is why only Herwig\xx and not Pythia or Sherpa have such a significant
population of low-momentum quark ``pollution''. This deserves further
investigation which would not be appropriate here, both investigating the
effects of various Herwig\xx model features and a more nuanced treatment of
$q$--$\gamma$ clustering in the QCD-aware algorithm.

In the absence of strong empirical motivations to choose the \akt or
\kt-reclustered labelling schemes, the \kt measure remains the most obvious
choice due to its theoretical links to QCD (and QED) emission kinematics in the
Sudakov regime, and because where scheme-dependent anomalies are seen, they
appear to be more prevalent in the \akt and reclustered-\kt schemes.

\subsection{Dependence on parton shower IR-cutoff and multi-parton interactions}
\label{sec:psmpisysts}


In this section we consider two possible systematic effects in the configuration
of the MC generator supplying events for jet finding and labelling: the choice
of parton shower cutoff scale, and the impact of multi-parton interactions. In
both cases the process of gluon splitting to quarks is the most dangerous to the
stability of the jet labelling, since a wide angle soft splitting may not be
clustered back together (without a distance modification \textit{\`a la}
flavour-\kt) in this early clustering phase, leaving two lone quark labels free
to contaminate other gluon pseudojets.

\subsubsection{Parton shower IR-cutoff}
\label{sec:pssysts}

In principle, the QCD-aware labelling technique can be passed a partonic final
state at any stage of its evolution and the results should remain largely
invariant, since the low-\pt evolution should be safely reversed in the
clustering. To test this hypothesis we have run Pythia\,8 with both a normal
($\mathcal{O}(1\;\GeV)$) parton shower cutoff, and one in which that cutoff has
been greatly raised to the jet \pt threshold of 30\;\GeV. The effect of this is
seen in Figure~\ref{fig:syscmplight}, and acts both as a very
conservative estimate to the algorithm's sensitivity to the IR region
of shower evolution (where different MC generators may use different
cutoff tunes) and an indication of the applicability of the algorithm
to appropriately defined fixed-order simulations.

From these plots, comparing the blue raised-cutoff line to the red Pythia\,8
default configuration, it can be seen that the low-\pt splittings in the default
configuration increase the rate of light-quark jet labels, while having little impact on
the gluon-label rate.  As well as in the overall normalisations, this effect can
particularly be seen in the high ``shoulder'' in the bottom-right quark-label
\pt-resolution observable, which is most prominent for the red default
setup. This suggests that the relative ease of $q/g$ contamination leads to
``wrong'' light-quark labels being derived from gluon splittings to light quarks
in the high multiplicity of low-\pt shower branchings, and hints that an
improved matching requirement might prefer to label with the highest-\pt label
jet close to the particle jet axis in order to reduce this ``shoulder'' of
mismatched too-low-\pt labels.

\subsubsection{Multi-parton interactions (MPI)}
\label{sec:mpisysts}

Multi-parton scattering poses a potentially lethal threat to an algorithm
starting from final state partons, because of the intrinsic assumption that at
least a \kt-based clustering will be able to reverse the order of QCD
splittings. Typically we are only interested (as far as possible) in the hardest
partonic interaction, but MPI overlays the partonic final state of that
interaction with those from secondary partonic scatters. As a result of this
incoherent overlay of distinct partonic final states, the geometrically optimal
partonic clusterings may be between partons evolved from different hard
processes, which (at least in model terms) are unrelated. This is expected to be
particularly problematic for MPI \emph{quarks}, since unlike gluons (or photons)
they can lead to a reassignment of ``true'' gluon jet flavour labels.

\begin{figure}[tp]
  \centering
  \img[0.48]{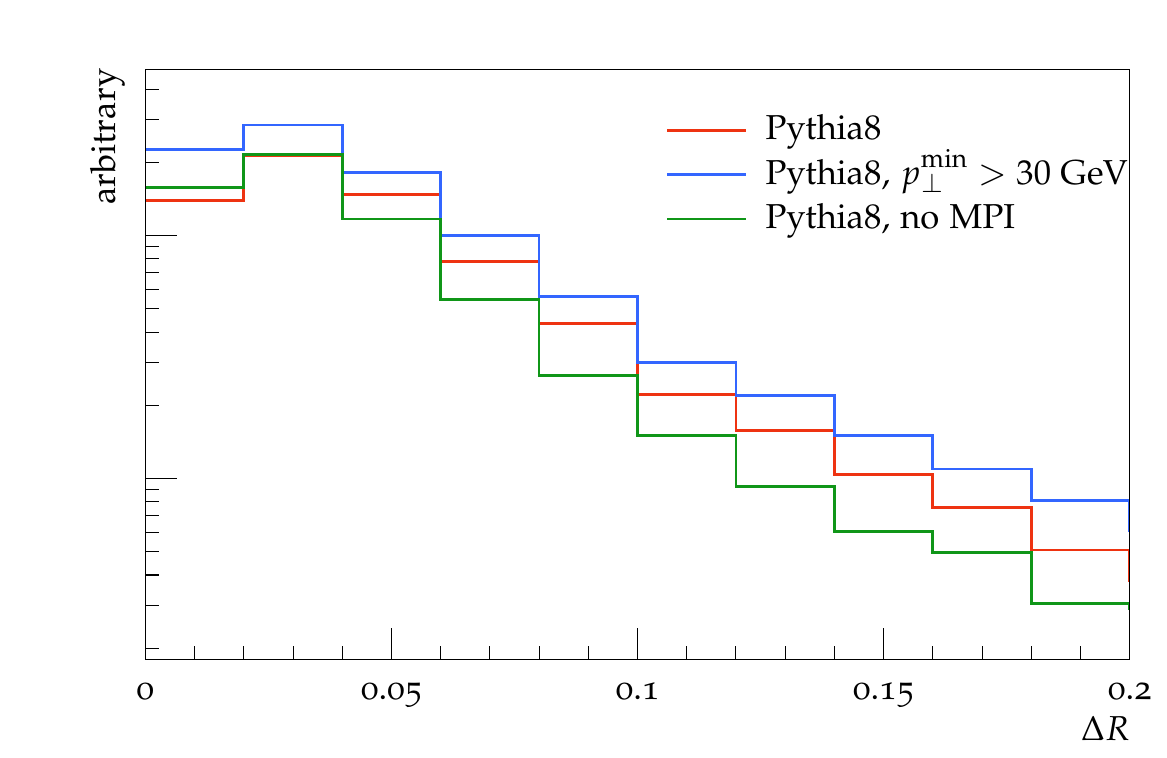} \quad \img[0.48]{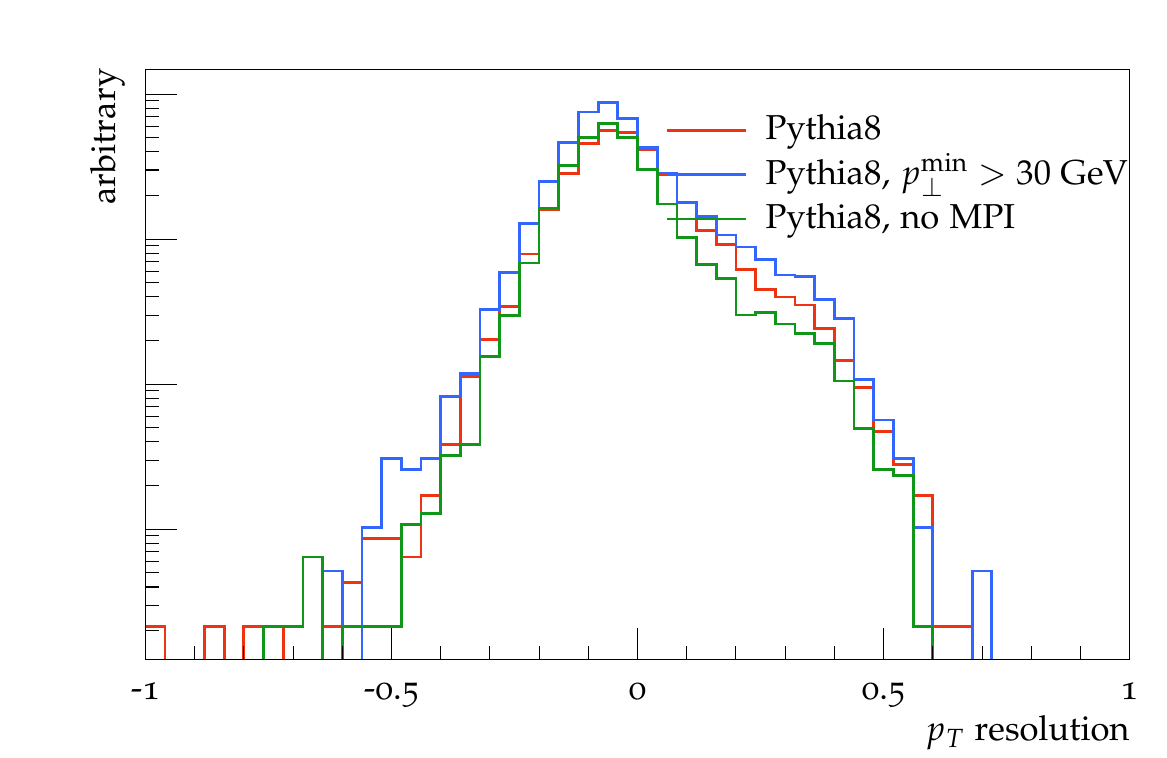}\\
  \img[0.48]{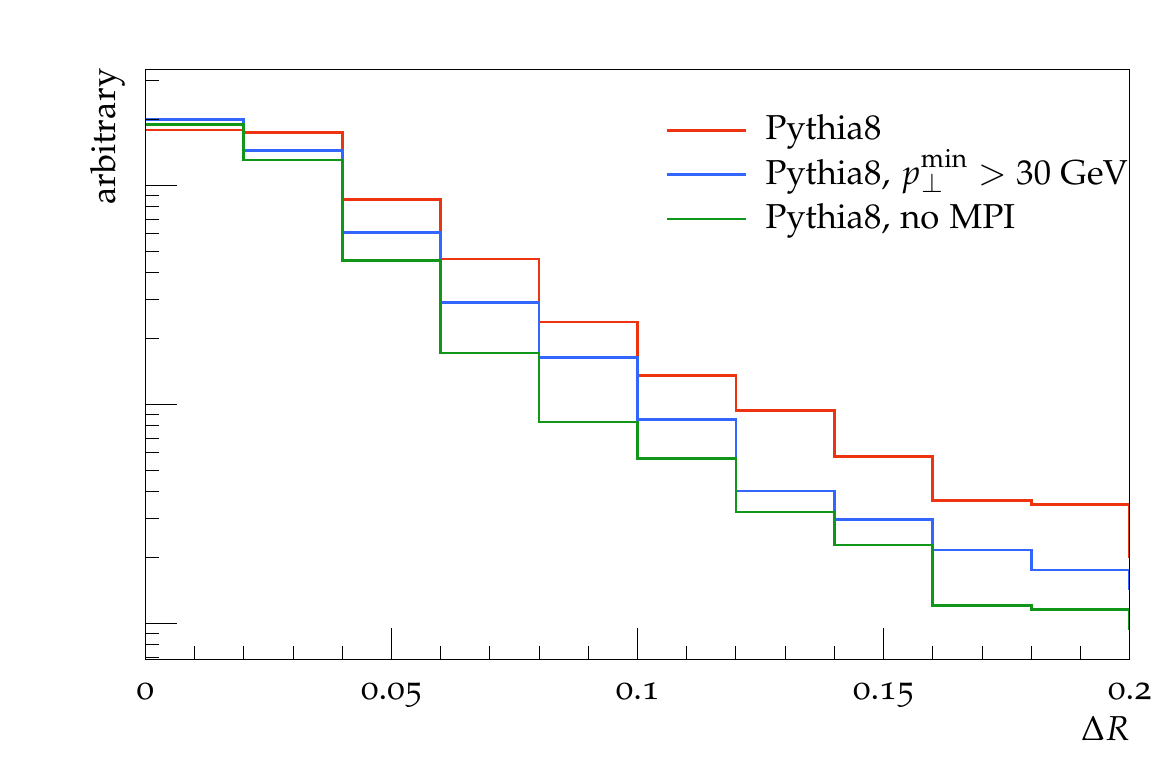} \quad \img[0.48]{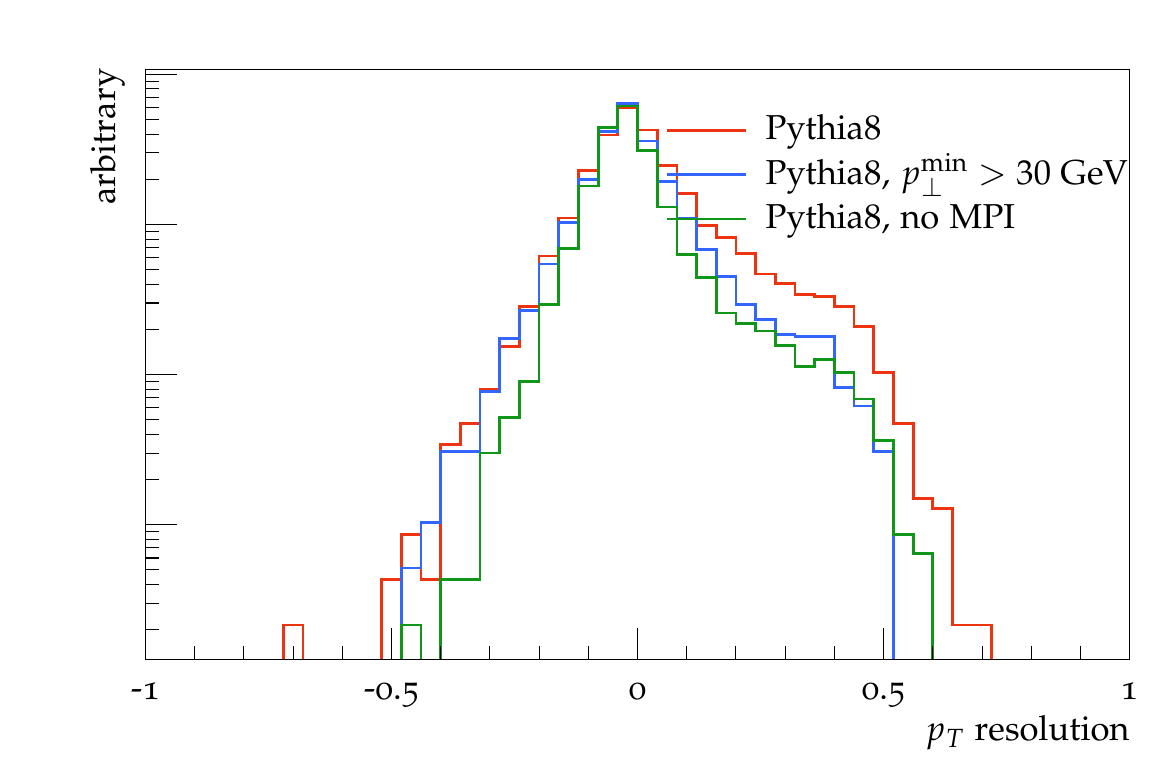}
  \caption{Light tagging dependence on event simulation systematics:
    inclusive jet events, with gluon-labelled jets on the top row, and
    light-quark-labelled jets on the bottom.}
  \label{fig:syscmplight}
\end{figure}

All the performance plots shown so far have had MPI modelling enabled, and in
the label ratio discrepancies seen in \Fref{sec:ratios} -- particularly for
Herwig\xx with respect to the other generators -- there are hints that MPI
overlay could be responsible. The $gg \to
gg$ process dominates the low-$x$ MPI processes
but shower evolution of the MPI partons would be expected to produce some
unpaired gluon splittings with a resulting effect. In \Fref{fig:syscmplight} we
also compare Pythia\,8 with and without MPI modelling enabled, to gauge the
magnitude of the effect.

It is clear that the green no-MPI configuration gives substantially better
$\Delta{R}$ and \pt agreement between the particle jets and their labels, as
well as slightly reducing the overall normalisation -- both effects are seen in
both the gluon and quark distributions. Resolution improvements were also seen,
in the unshown charm and bottom performance measures.  This normalisation
change, however, does not relate to an increased rate of unlabelled jets, but
just the lower total number of particle jets in a no-MPI generator configuration
(this was checked separately and is not evident from the displayed plots). While
the removal of MPI produces narrower distributions than default in all cases,
the largest effect is again seen on the upper side of the light quark
\pt-resolution observable, where adding soft MPI emissions enhances the
\pt-mismatched ``shoulder'' to the same extent as soft shower splittings did in
the previous section.

\begin{table}[tp]
  \centering
  \begin{tabular}{ll@{\qquad}r@{\qquad}rr}
    \toprule
            &           & Jets  & \multicolumn{2}{c}{$\gamma + \text{jet}$} \\
    Scheme  & Generator & $q/g$ & $\gamma/g$ & $q/g$ \\
    \midrule
    Max-\pt & Pythia\,8 & 0.39  & 15.4       &  9.5 \\
            & Herwig\xx  & 0.33  & 18.3       & 11.4 \\
            & Sherpa    & 0.57  & 13.4       &  7.0 \\
    \addlinespace
    \kt     & Pythia\,8 & 0.65  & 11.8       &  7.6 \\
            & Herwig\xx  & 0.68  & 11.2       &  8.0 \\
            & Sherpa    & 0.73  & 13.0       &  7.0 \\
    \addlinespace
    \akt    & Pythia\,8 & 0.65  & 11.7       &  7.6 \\
            & Herwig\xx  & 0.93  & 11.0       &  8.1 \\
            & Sherpa    & 0.74  & 12.9       &  7.0 \\
    \addlinespace
Reclustered & Pythia\,8 & 0.64  & 11.5       &  7.5 \\
            & Herwig\xx  & 0.80  & 11.0       &  8.2 \\
            & Sherpa    & 0.73  & 12.7       &  6.9 \\
    \bottomrule
  \end{tabular}
  \caption{Jet label ratios for the combined sample of leading and subleading jets
    constructed in inclusive jet and $\gamma+\text{jet}$ simulated events, for MC
    generators with MPI modelling disabled.}
  \label{tab:ratiosnompi}
\end{table}


%
%
%


In Table~\ref{tab:ratiosnompi} we show the same jet label ratios as computed
before, but now with MPI disabled for all three generators. The difference is
striking: the Pythia8 and Sherpa $q/g$ ratios in jet events are little changed,
but Herwig\xx's ``inverted'' ratio is now shifted to agree with the others; and
in $\gamma+\text{jet}$ events the high Herwig\xx gluon rate is now gone, bringing
its ratios in line with the other generators (and with very stable,
cross-generator values in the QCD-aware scheme). MPI is clearly a very significant
problem for post-fragmentation truth-jet labelling algorithms to address.


\FloatBarrier

\subsection{Dependence on higher-order ME parton production}
\label{sec:mesysts}


In this section we make a final study of the robustness of the QCD-aware
labelling scheme, considering variations in the maximum number of matrix element
final-state partons for QCD jet events. This study was performed using the
Sherpa event generator, configured in three separate runs to use a lowest-order
$2 \to 2$ QCD scattering matrix element as in the previous studies, as well as
higher-order tree-level $2 \to 3$ and $2 \to 4$ MEs.

\begin{figure}[tp]
  \centering
  \img[0.48]{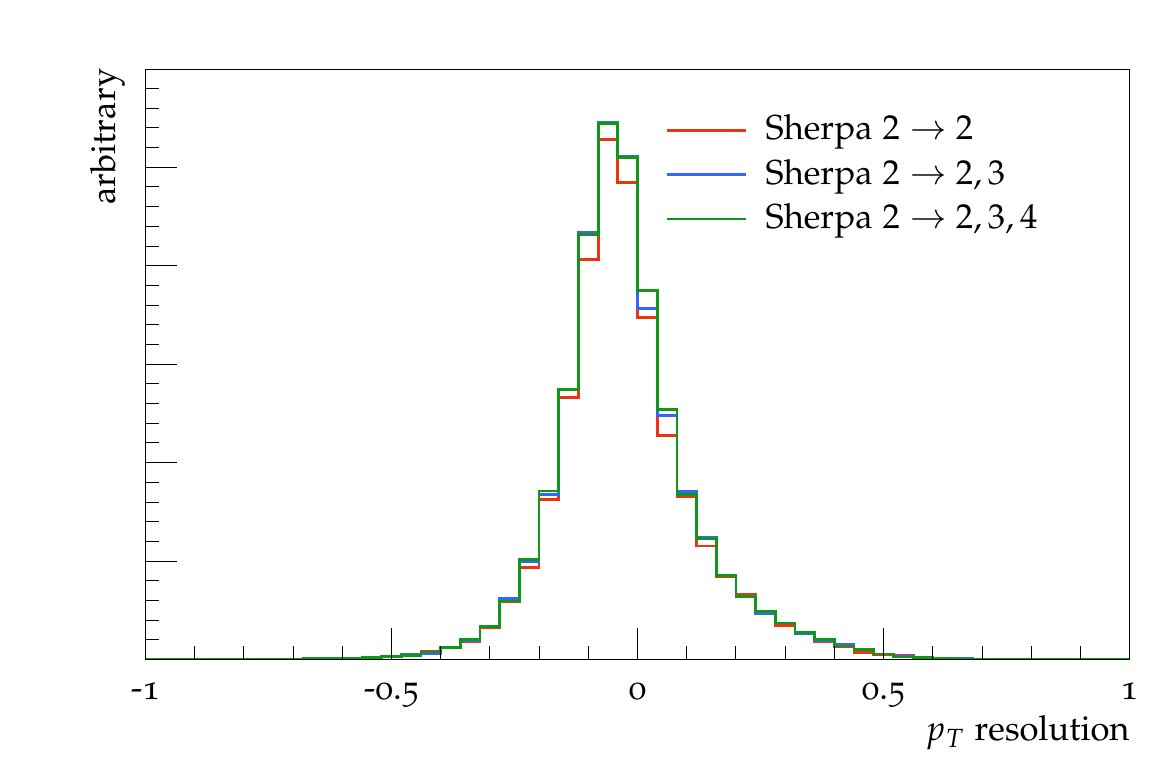} \quad \img[0.48]{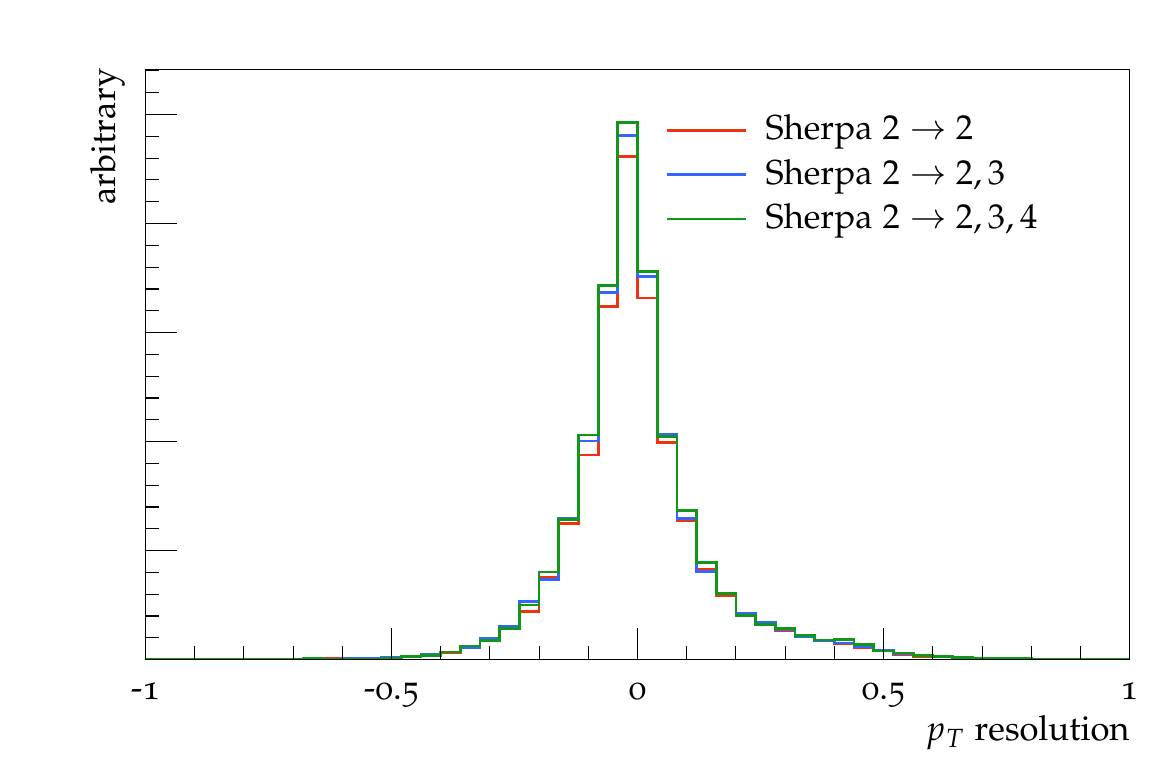}\\
  \img[0.48]{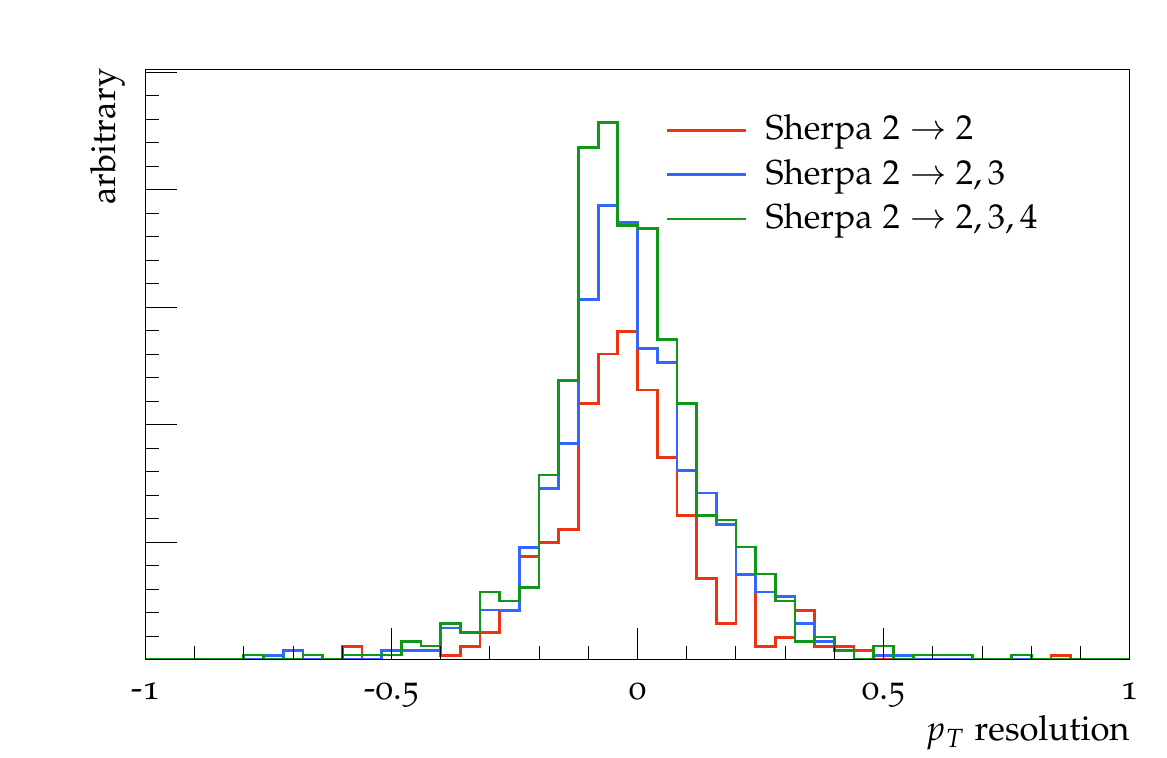} \quad \img[0.48]{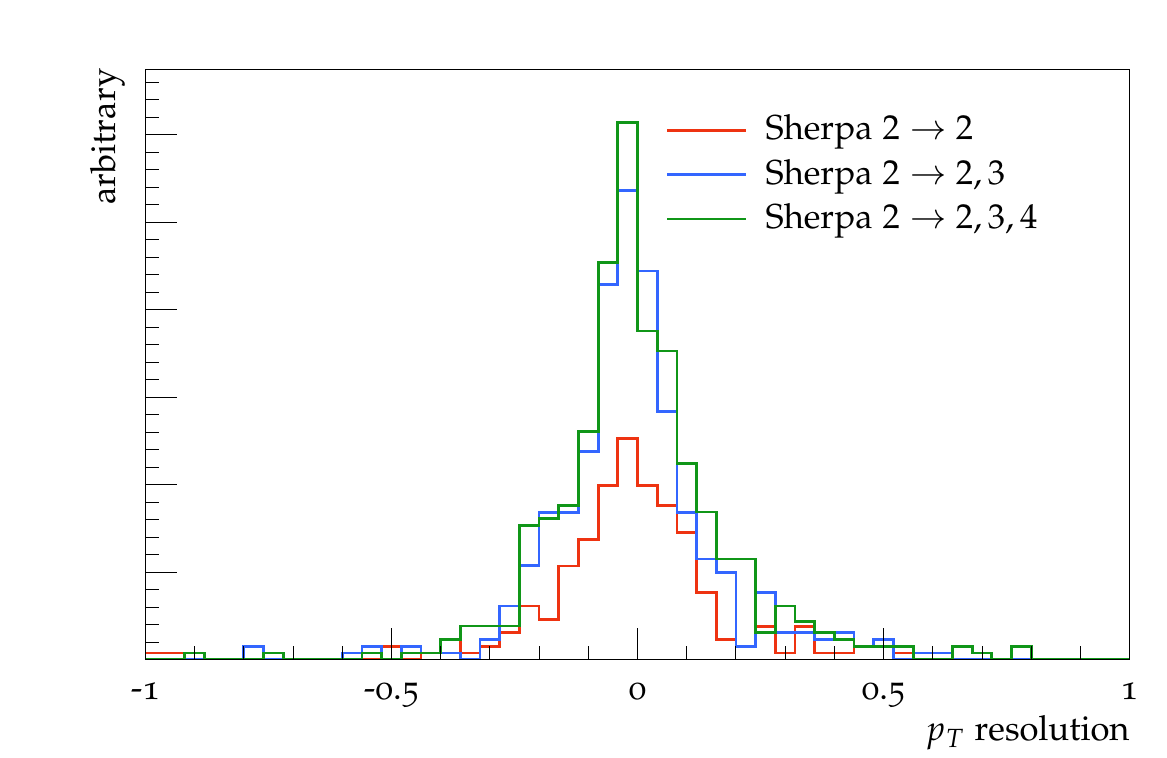}
  \caption{Jet labelling performance dependence on maximum number of
      ME partons in jet events for gluon- (left) and light
      quark-labelled jets (right). The top row shows all jets
  inclusively, the bottom only for the 3rd jet.}
  \label{fig:mecmpdpt}
\end{figure}

The dominance of the lowest-order hard process configuration is illustrated by
the stability of the cross-section which rose only by 0.1\% from the $2 \to 2$
value of 3.073\;\mb with the addition of diagrams with up to two extra
final-state partons. The inclusive jet plots in \Fref{fig:mecmpdpt} show a
larger effect, since the total number of jets to pass the 25\;\GeV analysis \pt
cut increased by 4.5\% and 5.7\% over the 2-parton configuration for the 3- and
4-parton samples respectively. But the stability of the performance measures is
still notable.

\begin{table}[t]
  \centering
  \begin{tabular}{lrrrrr}
    \toprule
    ME & $N_\mathrm{j3}/N_\mathrm{j3}^{2\to2}$ & Gluon frac. & Light quark frac. & Light parton frac. & Unlabelled frac.\\
    \midrule
    $2 \to 2$ & 1.00 & 62.7\% & 27.0\% & 89.6\% & 2.3\% \\
    $2 \to 3$ & 1.59 & 56.4\% & 31.4\% & 88.3\% & 2.9\% \\
    $2 \to 4$ & 1.79 & 58.3\% & 31.9\% & 90.2\% & 2.6\% \\
    \bottomrule
  \end{tabular}
  \caption{Comparison of 3rd jet rates and flavour label fractions between Sherpa calculation with three different maximal ME final-state multiplicities.}
  \label{tab:mecmpjet3ratios}
\end{table}

We see larger differences if we focus on the 3rd-hardest jet in the event, which
in the $2 \to 2$ ME configuration should virtually never directly correspond to
an ME parton. The usual labelling performance measures are shown for
this in the bottom row of \Fref{fig:mecmpdpt}.
It is not simple to interpret these plots because the
dominant effect is the change in normalisation due to the increased total number
of 3rd jets as the matrix elements include more hard corrections: the increases
are 59.1\% and 79.3\% above the $2 \to 2$ configuration for the 3- and 4-parton
MEs respectively. If we instead look at ratios of light quark and gluon labels
to the total, shown in \Fref{tab:mecmpjet3ratios}, then stability is again
evident: gluon labels account for 58--63\% and light quarks for 27--32\% of 3rd
jets despite the large normalisation changes. The total number of jets labelled
as either gluon or light quark remains between 88\% and 90\% of the total, and
the fraction of unlabelled jets is also stable between 2--3\%.  No asymmetry or
change of distribution widths is observed with the changes of ME multiplicity.

We note for clarity that it is not the case that a perfect labelling algorithm
would give perfectly stable results in these tests -- after all, the physics is
being improved with each extra parton, changing the jet kinematics and
potentially the flavour balance. But it does suggest both algorithmic robustness
and that the light-flavour weighting (if not the kinematics) of the Sherpa
parton shower splitting functions is well-matched to the explicit ME
calculations.



\section{Conclusions and proposals}


Just as there is no absolute definition of what constitutes a hadronic jet,
there is no absolute way to assign a flavour label to it. But, as for jet
finding, there are differences in the algorithms which are used as operational
definitions, and not all algorithms are equal.

In this spirit, we have described and characterised the performance of the
QCD-aware algorithm for truth jet partonic labelling. This algorithm is based on
restricting flavour combinations to those permitted within QCD and QED, and on
final state partonic inputs defined in the lab frame. It offers a
theory-motivated labelling approach portable between all the major families of
parton shower event generators, and shows fairly low sensitivity to calculation
artefacts such as parton evolution cutoffs and ME order. Comparable labelling
performance was seen between the generator families, across a range of hard
process types with multi-parton interactions (MPI) disabled: the ratios of jet
label rates from the three generators were in good agreement with each other,
and the dominant labels agreed with the prediction from fixed-order
cross-sections. A lack of strong dependence on distance measure was observed,
due to the constraining effects of the flavour combination rules. Hence, while
``just another algorithm'', we contest that the QCD-aware approach is better
theoretically motivated than existing approaches and portable between MC shower
generators in the absence of MPI, e.g. in $e^+e^-$ events.

However, the labelling performance \emph{is} affected by MPI, which introduces a
``shoulder'' structure several times higher than the no-MPI rate into the tail
of the labelling jet \pt resolution, and induces dramatic changes in the jet
label ratios computed for Herwig\xx. This is a problem which must be addressed
for use of this algorithm in fully-dressed hadron collider event simulations,
and studies are underway to improve the robustness to MPI contamination. The
exact reason for Herwig\xx's extreme labelling susceptibility to MPI effects is
not yet known. Changing the jet--label matching criteria to reject matches from
label partons much softer than the particle jet will address some issues, and it
is possible that suppressing flavour-changing clusterings with very soft quarks
-- as in the flavour-\kt clustering algorithm -- may be an effective MPI
rejection heuristic.

The code implementation of this algorithm, available in the FastJet contrib
repository as \kbd{QCDAware}, allows for flexible ``hybrid'' approaches such as
QCD-aware \kt reclustering of labelling partons selected by association to \akt
particle jets, for increased consistency with LHC experiment procedures.


It remains to be seen whether there is substance to hints that the labelling
results may slightly overestimate the rate of quark jets. This is quite
conceivable since pseudojet quark labels, once acquired, are harder to lose than
gluon labels are: overlay of quarks from MPI can easily switch the label of an
aligned hard-process gluon jet.
This is a clear area for further investigation, in which
the most obvious step is to introduce flavour-\kt style extra weighting for
quark \& gluon kinematics, as well as including the effects of different colour
and EM charges, and the relative (running) strengths of the QED and QCD
couplings. These extensions provide a clear motivation to focus on the \kt
measure as the canonical distance choice for QCD-aware labelling.

Finally, we note that the inclusion of leptons and photons in the QCD-aware
combination rules provides an attractive way to define truth-level dressed leptons and
isolated photons in addition to jets, without overlaps or false distinctions
between e.g. hard and ``fragmentation'' photons. Thus it may offer an appealing
alternative truth object definition scheme to those already on the
market~\cite{Buckley:1700541}. There is perhaps also the possibility to extend
the QCD-aware scheme with further processes such as EW resonance decays which
(like for partons) may not be reliably encoded in shower generator records:
e.g. $\bar{q_u} q_d' \to W^-$ and $W b \to t$ clusterings could be added with
appropriate clustering weights.
However, as the kinematics of these processes are non-Sudakov, there is no
guarantee that they would work optimally, or even at all.

\section*{Acknowledgements}

Our thanks to Donatas Zaripovas for his early work on testing the behaviour of
the QCD-aware algorithm, to Gavin Salam for clarifying our questions about the
impact and necessity of flavour-\kt distance measure modifications, and to Jesse
Thaler for feedback on a draft version of this paper. And of course to the
FastJet team for providing such a useful tool and their ``contrib'' repository
for collecting community contributions such as the code described here.

\bibliographystyle{atlasnote}
\bibliography{refs}

\clearpage
\appendix
\section{Further performance plots}
\label{app:extraplots}

\begin{figure}[hbp]
  \centering
  \img[0.48]{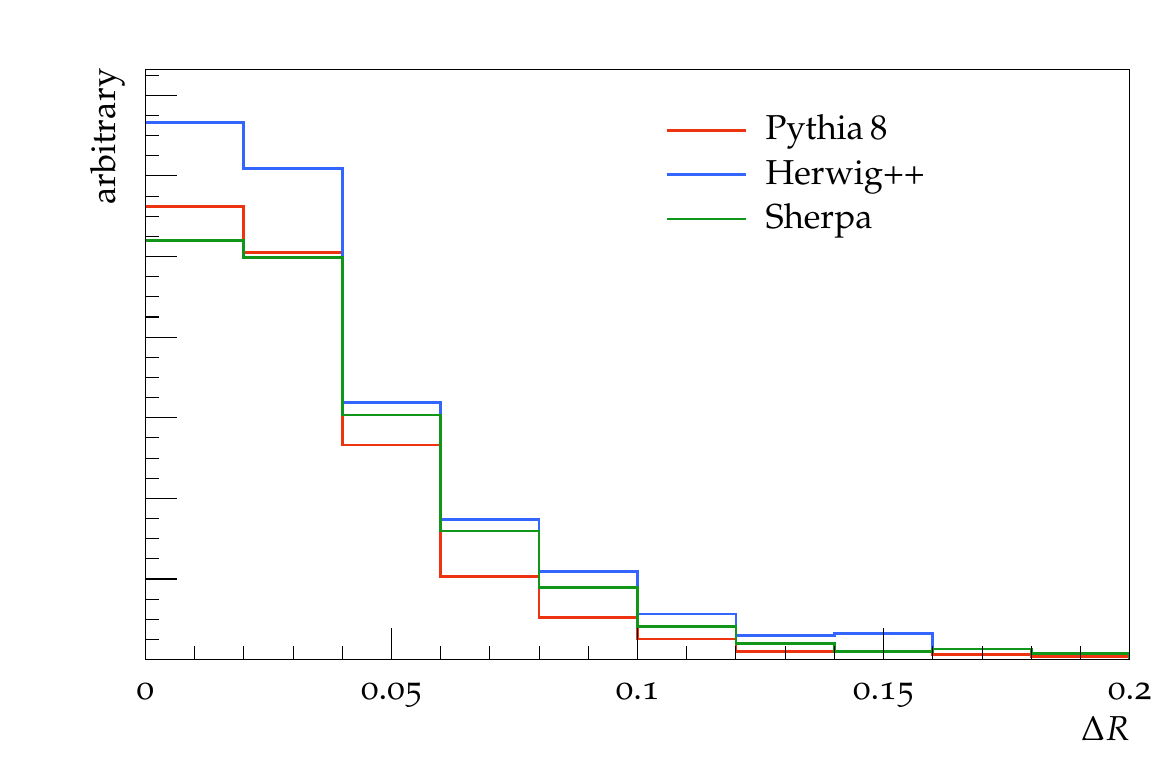} \quad \img[0.48]{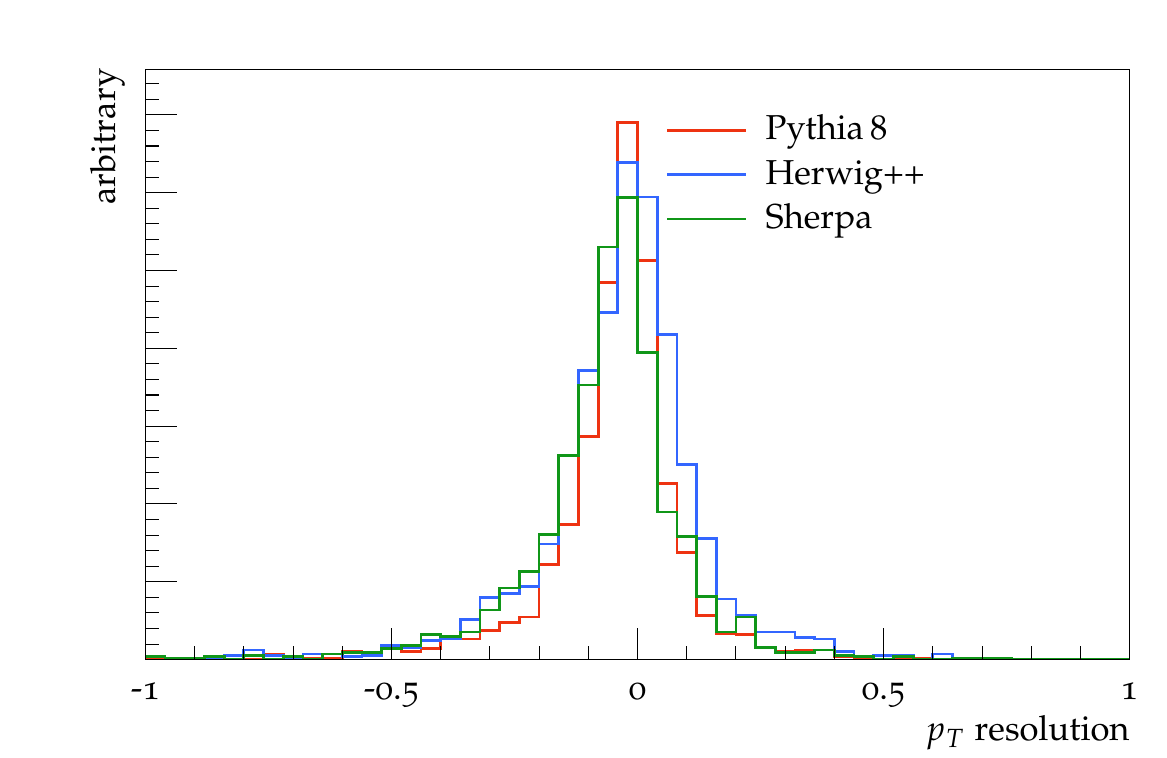}\\
  \img[0.48]{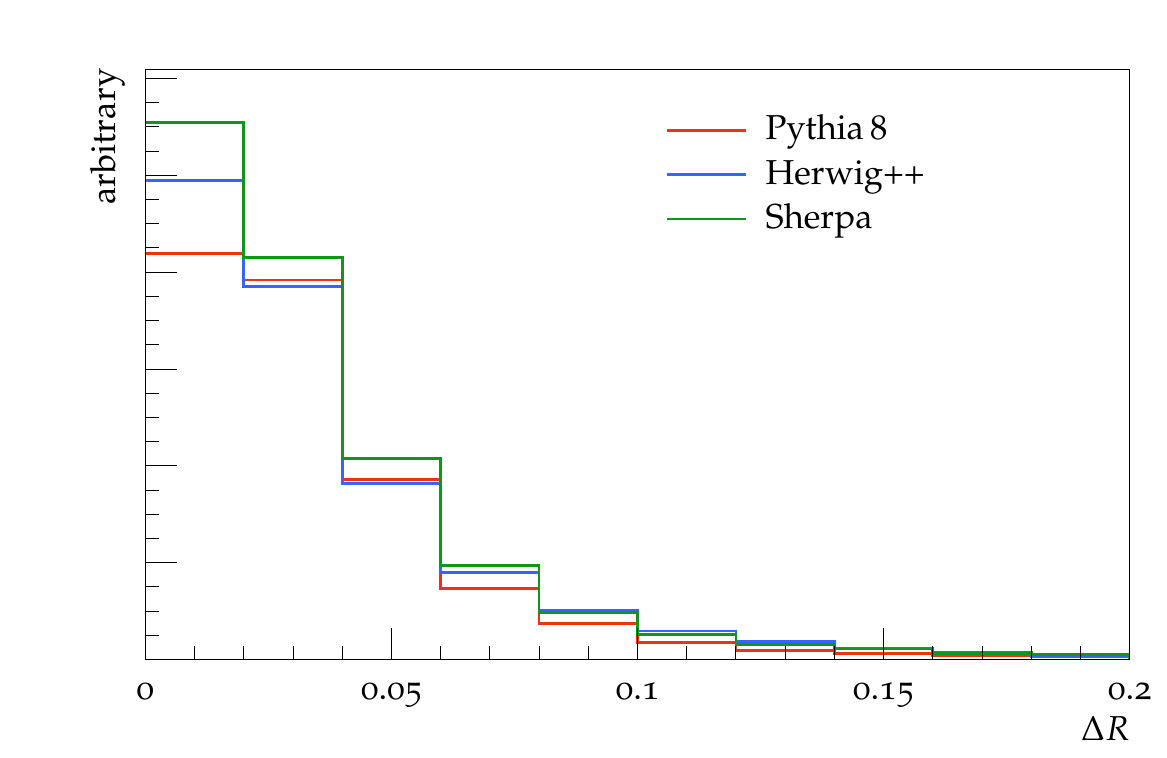} \quad \img[0.48]{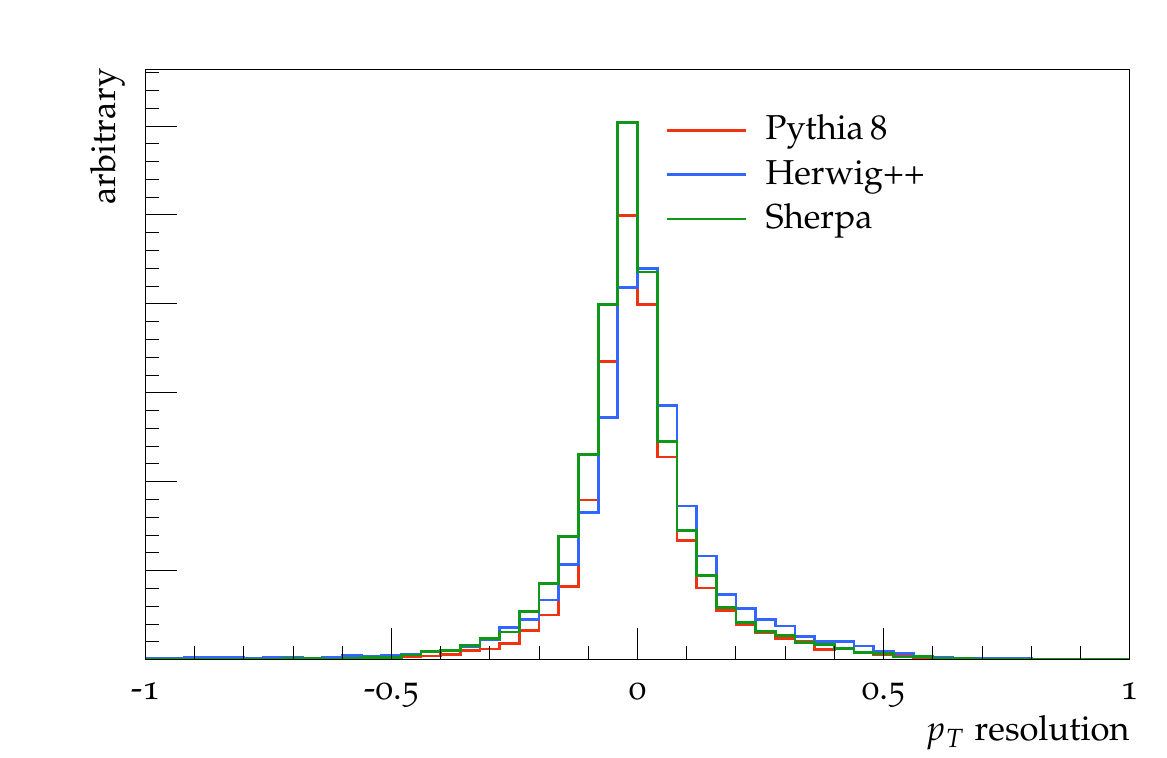}
  \caption{Heavy tagging performance/comparisons: $\gamma + \text{jet}$ events, with bottom-labelled
    jets on the top row, and charm-labelled jets on the bottom row.}
 \label{fig:gammajetcmpheavy}
\end{figure}

\begin{figure}[tp]
  \centering
  \img[0.48]{figs/jets2/Inclusive_Gluon_Kt_Dr.pdf} \quad \img[0.48]{figs/jets2/Inclusive_Light_Kt_Dr.pdf}\\
  \img[0.48]{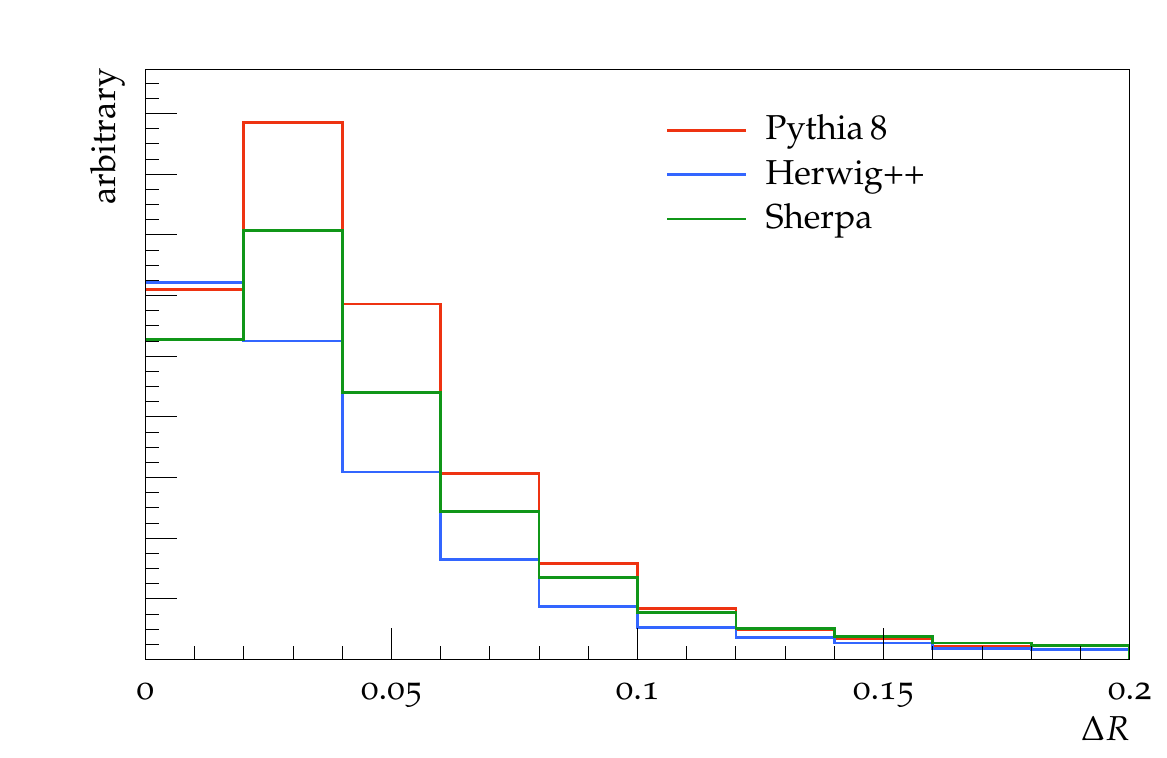} \quad \img[0.48]{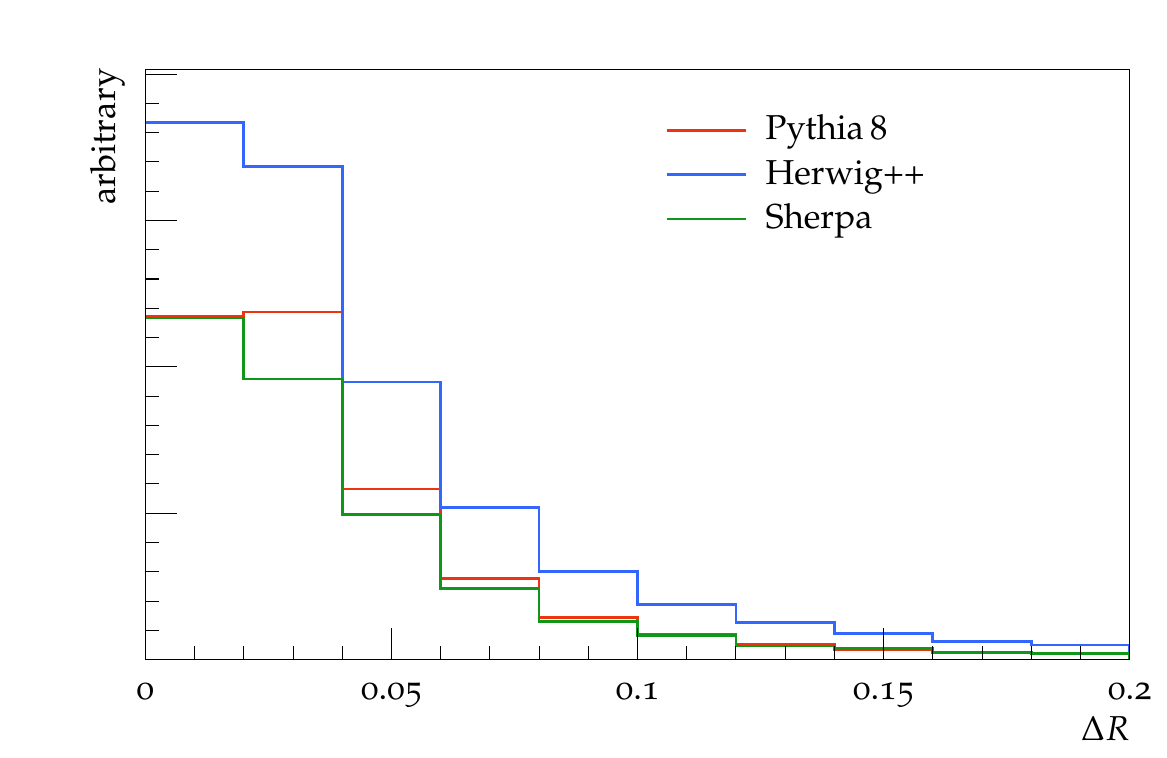}\\
  \img[0.48]{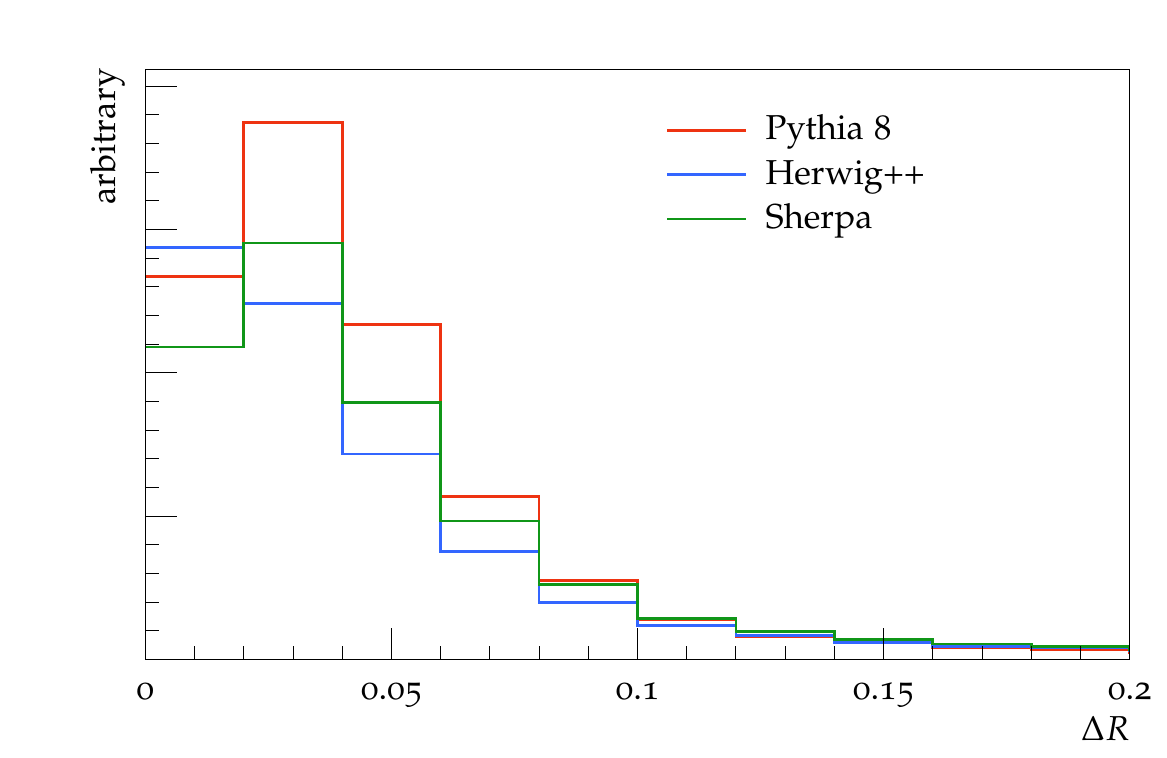} \quad \img[0.48]{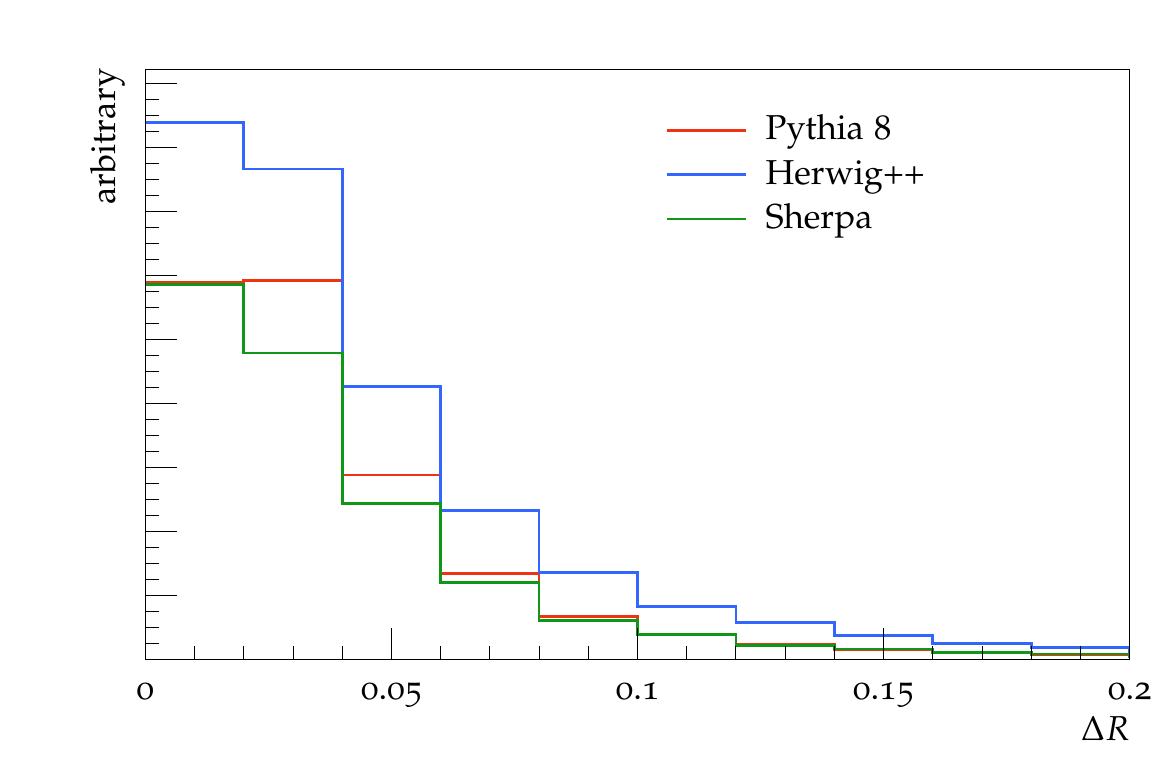}\\
  \caption{$\Delta{R}$ performance comparisons for gluon- (left) and
      light quark-labelled jets (right) in inclusive jet events, showing \kt
      labeled jets on the top row, anti-\kt in the middle, and
  \kt-reclustered on the bottom.}
  \label{fig:labelcmpdr}
\end{figure}

\begin{figure}[tp]
  \centering
  \img[0.48]{figs/jets2me/Inclusive_Gluon_Kt_Dpt.pdf} \quad \img[0.48]{figs/jets2me/Inclusive_Light_Kt_Dpt.pdf}\\
  \img[0.48]{figs/jets2me/Jet2_Gluon_Kt_Dpt.pdf} \quad \img[0.48]{figs/jets2me/Jet2_Light_Kt_Dpt.pdf}
  \caption{Jet labelling dependence on maximum number of ME partons in
      jet events for gluon- (left) and light quark-labelled jets
      (right).
      The top row shows all jets inclusively, the bottom only for the
      3rd jet.}
  \label{fig:mecmpdr}
\end{figure}

\end{document}